\documentclass[12pt]{article}

\def\Tr{\rm Tr\ }
\def\mt{\tilde{m}}
\def\varphit{\tilde{\varphi}}
\def\phit{\tilde{\phi}}

\newcommand{\ie}{{\em i.e.}\ }
\newcommand{\eg}{{\em e.g.}\ }

\def\fracs#1#2{\textstyle\frac #1#2}
\def\Dslash{\not{\hbox{\kern-4pt $D$}}}
\def\Bslash{\not{\hbox{\kern-4pt $B$}}}
\def\dslash{\not{\hbox{\kern-2pt $\partial$}}}
\def\gslash{\not{\hbox{\kern-2pt $\gamma$}}}
\def\pslash{\not{\hbox{\kern-2pt $p$}}}
\def\psislash{\not{\hbox{\kern-2pt $\psi$}}}
\newcommand{\be}{\begin{equation}}
\newcommand{\ee}{\end{equation}}
\newcommand{\ba}{\begin{eqnarray}}
\newcommand{\ea}{\end{eqnarray}}

\def\sqr#1#2{{\vcenter{\vbox{\hrule height.#2pt
         \hbox{\vrule width.#2pt height#1pt \kern#1pt
            \vrule width.#2pt}
         \hrule height.#2pt}}}}
\def\square{\mathop{\mathchoice\sqr56\sqr56\sqr{3.75}4\sqr34\,}\nolimits}

\def\La{\Lambda}

\def\Si{{\Sigma}}
\def\eql{{~=~}}

\def\Si{{\Sigma}}
\def\al{\alpha}

\def\cN{{\cal N}}
\def\cV{{\cal V}}

\begin{document}

\setlength{\baselineskip}{14pt} 
\setlength{\parskip}{1.35ex}
\setlength{\parindent}{0em}
\noindent

\thispagestyle{empty}
{\flushright{\small CITUSC/00-049, DAMTP-2000-106, MIT-CTP-3023, \\
NSF-ITP-00-105, ROM2F/2000/29B\\ hep-th/0009156\\}}

\vspace{.3in}
\begin{center}\Large {\bf 
Anatomy of Two Holographic \\ Renormalization Group Flows}
\end{center}

\begin{center}
{\large Massimo Bianchi\footnote{Dipartimento di Fisica and I.N.F.N.,
Universit\`a degli studi di Roma ``Tor Vergata'', Via della Ricerca
Scientifica, 00133 Rome, ITALY, and Department of Applied Mathematics
and Theoretical Physics, University of Cambridge - CMS, Wilberforce
Road, Cambridge, England CB3 0WA; bianchi@roma2.infn.it}, Oliver
DeWolfe\footnote{Center for Theoretical Physics, Massachusetts
Institute of Technology, Cambridge, MA 02139, U.S.A., and Institute
for Theoretical Physics, University of California, Santa Barbara, CA
90136, U.S.A;\\ odewolfe@itp.ucsb.edu}, \\ Daniel
Z. Freedman\footnote{Department of Mathematics and Center for
Theoretical Physics, Massachusetts Institute of Technology, Cambridge,
MA 02139, U.S.A., and Centre Emile Borel, UMS 839 IHP,
(CNRS/UPMC)--Paris, FRANCE; dzf@math.mit.edu} and Krzysztof
Pilch\footnote{CIT-USC Center for Theoretical Physics and Department
of Physics and Astronomy, University of Southern California, Los
Angeles, CA 90089-0484, U.S.A; pilch@usc.edu }}
\end{center}

\begin{center}September 2000\end{center}

\begin{abstract}
We derive and solve a subset of the fluctuation equations about two
domain wall solutions of $D=5$, ${\cal N}=8$ gauged supergravity.  One
solution is dual to $D=4$, ${\cal N}=4$ SYM theory perturbed by an
${\cal N}=1$, $SO(3)$-invariant mass term and the other to a Coulomb
branch deformation.  In the first case we study all $SO(3)$-singlet
fields.  These are assembled into bulk multiplets dual to the stress
tensor multiplet and to the ${\cal N}=1$ chiral multiplets ${\Tr}
\Phi^2$ and ${\Tr} W^2$, the former playing the role of anomaly
multiplet.  Each of these three multiplets has a distinct spectrum of
``glueball'' states.  This behavior is contrasted with the Coulomb
branch flow in which all fluctuations studied have a
continuous
spectrum above a common mass gap, and spontaneous breaking of
conformal symmetry is driven by a bulk vector multiplet.  $R$-symmetry
is preserved in the field theory, and correspondingly the bulk vector
is dual to a linear anomaly multiplet.  Generic features of the
fluctuation equations and solutions are emphasized.  For example, the
transverse traceless modes of all fields in the graviton multiplet can
be expressed in terms of an auxiliary massless scalar, and gauge
fields associated with $R$-symmetry have a universal effective mass.
\end{abstract}

\vspace{.1in}

\newpage
\section{Introduction}

It is quite remarkable that the $AdS$/CFT correspondence \cite{MAGOO}
describes the large $N$ strong coupling limit of 4D gauge theories
with both exact and broken conformal symmetry. There is by now a large
literature discussing the situation in which the renormalization group
flow of field theory couplings is, in principle, described by
supergravity; it is difficult to be complete, but a representative
sample of this work may be roughly divided into the categories of i)
early examples \cite{GPPZ2}-\cite{GPPZ1}, ii) explicit domain wall
solutions of $D=5$ and $D=10$ supergravities \cite{FGPW1}-\cite{BHLT},
and iii) holographic formulation of renormalization group equations
\cite{alvarez}-\cite{BGM}.

A holographic RG flow is in general described by a solution of a $D
\ge 5$ supergravity theory whose symmetry group is the 4D
Poincar\'e group or a supersymmetric extension thereof. The solutions of
linearized fluctuation equations in the bulk geometry can, in
principle, be used to calculate correlation functions of operators in
the dual boundary field theory. Our purpose is to derive and solve
these fluctuation equations in examples of the two main cases of
physical interest, the case where the bulk flow is dual to a massive
deformation of ${\cal N}=4$ Super Yang-Mills (SYM) theory \cite{GPPZ} and
the case where it is dual to a Coulomb branch vacuum of ${\cal N}=4$
SYM \cite{FGPW2,BS}.

Both bulk geometries were derived from the maximal gauged ${\cal N}=8,
D=5$ supergravity \cite{PPvN,GRWlett}. We select for study the
fluctuations of fields in the bulk graviton multiplet together with
other multiplets to which the graviton and its superpartners are
coupled. This allows us to explore the way in which conformal and
$R$-symmetries are broken in the boundary theory and to illuminate the
relation between bulk and boundary supersymmetry.  We also want to
contrast the behavior in the two situations of operator and Coulomb
branch deformations. The fluctuation equations initially extracted
from ${\cal N}=8$ supergravity are quite complicated, but we are able
to manipulate them into simpler ``universal'' forms which appear to be
more general than the specific flows from which they were obtained.
All fluctuation equations are effectively hypergeometric, and their
solutions give the essential features of the ``glueball spectra'' in
both flows we study.

Most of our attention will focus on the flow proposed by Girardello,
Petrini, Porrati and Zaffaroni (GPPZ) \cite{GPPZ}, where a kink
solution was found involving a scalar field $m(r)$ which turns on a
common mass for the three ${\cal N}=1$ chiral multiplets of ${\cal N}=4$
SYM theory. The massive theory flows toward pure ${\cal N}=1$ SYM at
large distance. The field theory interpretation of \cite{GPPZ} poses
several problems, and there may be inherent difficulties in a
5D approach in which curvature singularities are
endemic. Indeed very interesting non-singular descriptions of the
massive theory, based on 3- and 5-branes in $D=10$ Type IIB
supergravity have appeared \cite{PS,KlS,MN}.  Also recently, the
metric and dilaton/axion fields of the ``lift'' of the GPPZ flow to
ten dimensions were obtained \cite{PW}, suggesting new physical
features.  In any case, the GPPZ flow is a mathematically consistent
example of holography, and the boundary behavior of bulk fields agrees
with the field theory interpretation made in \cite{GPPZ}.  This should
be enough to illuminate the behavior of the correlators of important
operators such as the stress tensor and symmetry currents.

The breakdown of conformal symmetry in the massive field theory is
reflected in supergravity by mixing of the graviton trace $h_\mu^\mu$
with the scalar fluctuation $\tilde{m}$, as discussed in
\cite{DF,AFT}.  In addition to $h_{\mu\nu}$, the gravity multiplet
contains a symplectic Majorana gravitino $\psi_\mu^a, a= 1,2$, and a
$U(1)_R$ gauge field $B_\mu$.  Because of broken superconformal
symmetry and broken $R$-symmetry respectively, $\psi_\mu^a$ mixes with
a spin-1/2 field $\xi^a$, and $B_\mu$ mixes with a scalar
$\beta$. The fields $\xi^a,\tilde{m},\beta$ and two more scalars span
a bulk ${\cal N}=2$ hypermultiplet dual to the anomaly multiplet in
the field theory. There is a second, inert hypermultiplet containing
the dilaton dual to the operator ${\Tr} F^2 + \ldots$ and its SUSY
partners.

The graviton multiplet and these two hypermultiplets contain all bulk
fields which are singlets of the $SO(3)$ flavor symmetry preserved by
the GPPZ flow, and we will obtain and solve the fluctuation equations
for all these modes.  Although these singlet fields decouple from the
rest of the $\cN =8$ supergravity theory, their mutual interactions
are still rather intricate.  In particular, the dynamics of the 8 real
scalars determines a nonlinear $\sigma$-model on the quaternionic
manifold $G_{2(2)}/SO(4)$. Even at the level at which we work, which
is exact in 2 of the 8 scalars and bilinear in all other fields, the
extraction of the field equations for the singlet sector from the full
${\cal N}=8$ theory is a complex technical task that would be
difficult to carry out without extensive use of an algebraic
manipulation program.  We have used Mathematica to compute the full
scalar action for the $SO(3)$ singlet sector.  In particular, we have
obtained the exact potential for all 8 scalar fields; the calculation was
feasible using the so-called solvable parameterization of the scalar coset
(see, {\eg}, \cite{ADAFFT}).

The fluctuations of the transverse traceless components of $h_{ij}$,
the scalar $m$ and another scalar $\sigma$ have been obtained
previously \cite{DF,AFT,AGPZ}, and it is known that their
fluctuation equations can be transformed to hypergeometric form with 3
distinct hypergeometric solutions, whose asymptotics in turn determine
the discrete spectrum of dual field theory states.  We show explicitly
that all $SO(3)$-singlet boson and fermion fluctuations involve the
same three hypergeometric functions, corresponding to three distinct
glueball spectra for the dual operators, as we now summarize:

For the ${\cal N}= 1$ supercurrent multiplet ${\cal J}_{\alpha
\dot\alpha} = {\Tr} (W_{\alpha} \bar{W}_{\dot\alpha}+ \ldots )$ dual
to the transverse components of the bulk supergravity multiplet $\{
h_{\mu\nu}, \psi_{\mu}^{1,2}, B_{\mu} \}$, we have states with
momenta:
\be
\label{Jspec}
(pL)^2 = 4 (n+2)^{2} \qquad n=0,1,2, \ldots \,.
\ee
For the ${\cal N}= 1$ chiral anomaly multiplet ${\cal A} = { \Tr}
\sum_{i=1}^3 (\Phi_i^{2})$ dual to the active hypermultiplet $\{ \rho,\xi^{1,2}, \mt \}$:
\be
\label{Aspec}
(pL)^2 = 4 (n+1)(n+2) \qquad n=0,1,2,\ldots  \,.
\ee
For the ${\cal N}= 1$ chiral ``Lagrangian" multiplet ${\cal S} = {
\Tr} 
(W^\alpha W_\alpha + \ldots)$
 dual to the dilaton hypermultiplet $\{ \sigma, \xi^{3,4}, \tau \}$,
\be
\label{Sspec}
(pL)^2 = 4 n(n+3) \qquad n=0,1,2,\ldots \,,
\ee
including a zero-mass pole for the lowest component operators dual to
$\sigma$.

This pattern agrees with physical expectations, but it emerges in a
subtle way from the interwoven symmetries and dynamics of the bulk
supergravity theory. The dilaton and axion fields $\tau$ are treated
correctly for the first time; to do this forces us to confront the
complexity of the $G_{2(2)}/SO(4)$ coset.  In particular we find that
although the transverse traceless modes in the supergravity multiplet
can be conveniently written in terms of an auxiliary scalar field,
there is no physical scalar field with the same modes in the
theory\footnote{As discussed at the end of section~\ref{ScalarSubsec},
we believe that the dilaton was not treated correctly in
\cite{DZ}. Incorrect statements about a massless scalar fluctuation in
the GPPZ flow appear in \cite{DF,PZ}.}.

We then go on to explore the way spontaneously broken conformal
symmetry is realized in a supergravity background by examining the
fermion and vector sectors of a Coulomb branch flow, the ``$n=2$''
configuration corresponding to a disc of D3-branes found in
\cite{FGPW2,BS}. The active scalar, denoted by $\varphi(r)$, turns on
an expectation value for a real component of the scalar bilinear
${\Tr}(X^2)$ in the $20'$ representation of $SU(4)$.  This background
preserves $\cN = 4$ supersymmetry and $SU(2) \times SU(2) \times U(1)$
$R$-symmetry.  All field theory operators examined possess a continuous
spectrum above a common mass gap.

Examining the fluctuation equations, we find that the
gravitino/spin-1/2 sectors have an identical structure to
the GPPZ case, as was already known to occur with $h^\mu_\mu /
\varphit$.  The graviphoton remains massless, since the $U(1)$
$R$-current is conserved; however, it couples to the active scalar
background via a modified kinetic term.  This difference has its
origin in the fact that $\varphi$ is real and in a vector multiplet,
as contrasted with the complex $m$ which sits in a hypermultiplet.
This vector multiplet is naturally associated with a {\em linear}
anomaly multiplet in the dual field theory, as is known to arise in
non-conformal theories with preserved $R$-symmetry \cite{SW}.  Thus
two examples we consider demonstrate that the bulk multiplet containing the
active scalar is intimately linked to the type of anomaly multiplet
arising in the field theory.  Other recent work on the identification
between bulk and boundary supermultiplets can be found in \cite{BGMR}.
Finally, we show that all although massless vectors in the Coulomb
branch flow possess modified kinetic terms, their equations of motion
can be transformed to eliminate these in favor of the common mass term
$m_B^2 = - 2 A''$ with the same form as for the GPPZ graviphoton.

In section 2 we establish notation and review the metric, connections,
and Killing spinors for domain walls in 5 dimensions. The GPPZ and
Coulomb branch solutions are presented. In section 3 we discuss the
extraction of actions and transformation rules for the $SO(3)$-singlet
fields from the ${\cal N}=8$ theory and present results. In section 4
we show that the transverse traceless modes of all fields in the
graviton multiplet can be written in terms of an auxiliary massless
scalar field.  In section 5 we summarize and synthesize previous
results of \cite{DF,AFT} for the coupled $h^\mu_\mu / \mt$ sector and
present the 2-point function for the Coulomb branch case.  Section 6
is devoted to the decoupling and solution of fluctuation equations for
vector and scalar fields. The analogous discussion for the fermion
sector is given in section 7, in which we also verify the Bianchi
identity for the gravitino equation of motion.  The fermion and vector
fluctuations of the Coulomb branch flow are discussed in section 8,
and similarities and differences to the GPPZ case noted.  Although the
results of section 3 are the basis of the fluctuation equations solved
in later sections, these sections are largely self-contained and
can be understood without a detailed reading of Section 3.

The next step in this investigation will involve the systematic
application of results for bulk field fluctuations to obtain 2-point
correlation functions of the dual operators in the field theory and to
study physical implications. This question was addressed in
\cite{DF,AFT}, but there is still need to clarify the extraction of
field theory information from the supergravity fluctuations.  For
example, a procedure to obtain correlation functions of the stress
tensor which is invariant under gauge choices made in treating bulk
fluctuations is desirable. We hope to report on this soon.

\section{Review of GPPZ and Coulomb branch flows}

In this paper we shall be concerned with two backgrounds of
5D maximally supersymmetric gauged supergravity,
exemplifying the two distinct types of dual RG flows: the operator
deformation, which modifies the Lagrangian of the dual field theory,
and the Coulomb deformation, which only modifies the vacuum of the
theory.  In the examples we study there is one ``active'' scalar field
$\phi(r)$ which depends on the radial coordinate of the geometry.

In our conventions, the 5D Newton constant is such that $\kappa_5 = 2$
and the coupled gravity/scalar action is 
\begin{equation}
\label{GSaction}
S= \int \, d^5x  \sqrt{g} \, \left [-\frac{1}{4}R + 
 \frac{1}{2} g^{\mu\nu}\partial_\mu \phi \partial_\nu \phi
  -V(\phi) \right],
\end{equation}
where we use $(+----)$ signature. The background geometry always has
domain wall form
\begin{eqnarray}
\label{metric}
ds^2 &=& e^{2A(r)} \, (\eta_{ij} \, dx^i dx^j) - dr^2 \,.
\end{eqnarray}
The frames, nonvanishing Christoffel and spin connections, and
curvature tensors derived from the metric (\ref{metric}) are
\begin{eqnarray}
e^{\hat{k}} = e^{A} dx^k \,,    \quad  e^{\hat{r}} = dr \,, \quad 
\Gamma^r_{ij} = A' g_{ij} \,, \quad  \Gamma^i_{jr} = \Gamma^i_{rj} = A'
\delta^i_j \,, \quad
\omega_{j}^{\; \; \hat{k} \hat{r}} = - \omega_{j}^{ \; \; \hat{r} \hat{k}
} = - A' e^A \delta_j^{\hat{k}} \,, \\
R_{ijkl} =  (g_{ik} g_{jl} - g_{il} g_{jk}) A'^2 \,, \quad R_{irjr} = -
g_{ij} (A'^2 + A'')\,, \\ 
R_{ij} =  (4 A'^2 + A'') g_{ij} \,, \quad R_{rr} = - 4 A'^2 - 4 A'' \,, 
\quad R = 20 A'^2 + 8 A'' \,.
\end{eqnarray}
Here and throughout we use $i,j$ as 4D curved indices,
$\mu, \nu$ as 5D curved indices and $\hat{k}, \hat{l}$
as 4D Lorentz indices.

In the boundary limit $r\rightarrow \infty$ the geometry approaches
that of $AdS_5$, with asymptotic scale factor $A(r)\rightarrow r/L$,
while the active scalar, assumed dual to a field theory operator
${\cal O}_{\phi}$ of scale dimension $\Delta$, has the asymptotic form
$\phi(r) \sim e^{(\Delta-4)r}$ for operator perturbations and $\phi(r)
\sim e^{-\Delta r}$ for vacuum deformations.\footnote{For the case
$\Delta=2$, the scalings are $r e^{-2r}$ and $e^{-2r}$, respectively.}
Far from the boundary both background geometries we consider present curvature
singularities where $e^{2A(r)} \rightarrow 0$. Although this formally
indicates a breakdown of the supergravity description, we shall
proceed by choosing the solutions of all fluctuation equations to be
regular at the singularity, which is the established procedure.  The
singularities of both flows we consider satisfy the acceptability
criterion formulated by Gubser \cite{Gubs}.

The spacetime symmetry of the general domain wall is the 4D
Poincar\'e group. The special case when $A(r)=r/L$ and $\phi(r)=const$
corresponds to an exact $AdS_5$ geometry with an additional 5
isometries, dual to scale  and special conformal
transformations in the field theory.

The backgrounds we study are supersymmetric, which means that they
possess Killing spinors, zero modes of the spinor transformation
rules of the bulk supergravity theory \cite{Shus}.  To discuss these we recall
that the gauged ${\cal N}=8$ theory contains {\bf 8} symplectic Majorana
gravitino fields $\psi_{\mu a}$, with the $USp(8)$ index $a = 1, \ldots,
8$ raised and lowered by the symplectic metric $\Omega_{ab}$,   and
{\bf 48}
spinor fields $\chi_{abc}$ whose transformation rules are \cite{GRW}
\begin{eqnarray}
\label{psisusytrsf}
\delta \psi_{\mu a} &=& D_\mu \epsilon_a - \frac{g}{6}
W_{ab}(\phi) \gamma_\mu \epsilon^b \,, \\
\label{chisusytrsf}
\delta\chi_{abc} &=& \left[ \sqrt{2} \gamma^\mu P_{\mu abcd}(\phi) -
\frac{g}{\sqrt2}  A_{dabc}(\phi) \right] \epsilon^d \,,
\end{eqnarray}
where $g=2/L$ is the bulk gauge coupling constant, and the $USp(8)$ tensors
$W_{ab}$, $A_{abcd}$, and $P_{\mu abcd}$ are complicated functions of
the 42 scalar fields of the theory. Their expressions simplify vastly
after group theoretic analysis is used to truncate the equations of
motion of the theory to subsets of one or two scalars, a process first
done in the context of 5D RG flows in \cite{FGPW1}. On
such a scalar subset one examines the symplectic eigenvalue problem
for the symmetric matrix $W_{ab}$. Let $W(\phi)$ denote the eigenvalue
on a given two dimensional eigenspace spanned by a symplectic Majorana
pair $\epsilon_1, \epsilon_2$. On this subspace, the gravitino
transformation rule reduces to
\begin{equation}
\label{reducedkilling}
\delta \psi_{\mu a} = D_\mu \epsilon_a - \frac{g}{6}
W(\phi) \gamma_\mu \Omega_{ab} \, \epsilon^b \,.
\end{equation}
The first condition on Killing spinors is that they are zero modes of
(\ref{reducedkilling}).  This is achieved by imposing the first order
flow equation \cite{FGPW1},
\begin{eqnarray}
\label{gravkilling}
A'(r) = - \frac{g}{3} W(\phi(r)) \,,
\end{eqnarray}
which relates the scale factor and scalar field profile in the background.
Killing spinors then take the form
\begin{equation}
\label{SUSYkilling}
\epsilon = e^{A(r)/2} \eta^{(0)} \,, \quad \quad  
i\gamma^r \eta^{(0)}=\eta^{(0)} \,,
\end{equation}
where $\eta^{(0)}$ is a constant complex superposition 
$\eta^{(0)}= \eta^{(0)}_1 +i \eta^{(0)}_2$ 
on the original eigenspace; the chirality condition indicates
the Killing spinor contains a 4D constant Weyl spinor, so
it is 4D Poincar\'e supersymmetry which is naturally
associated with the domain wall.  In the complex notation the Killing
spinor (\ref{SUSYkilling}) satisfies
\begin{eqnarray}
\label{killingspin}
D_\mu \epsilon = i \frac{g}{6} W \gamma_\mu \epsilon \,.
\end{eqnarray}
The second condition on Killing spinors is that they are also
zero modes of (\ref{chisusytrsf}). This is assured if, on the same
eigenspace, the tensors $P_{\mu abcd}$ (linear in
derivatives of the active scalar $\phi(r)$) and $A_{abcd}$ simplify,
so that (\ref{chisusytrsf}) reduces to the flow equation \cite{FGPW1},
\begin{eqnarray}
\label{chikilling2}
\phi'(r) = \frac{g}{2} \,
\frac{\partial W(\phi)}{\partial \phi} \,.
\end{eqnarray} 
The flow equations (\ref{gravkilling}) and (\ref{chikilling2}) are
easily generalized to solutions with several active scalars and to the
case of a non-trivial $\sigma$-model metric. In particular
(\ref{chikilling2}) is just a gradient flow equation for the function
$W(\phi)$, which is called the superpotential for the active scalars
because it is related to the potential $V(\phi)$ on the reduced scalar
subspace of the flow by
\begin{equation} 
V(\phi) = g^2 \left[ \frac{1}{8} \left( \frac{\partial W(\phi)}{\partial \phi}
\right)^2 - \frac{1}{3} W(\phi)^2 \right] \,.
\end{equation}
The significant feature of the first order flow
equations is that any solution $\{ \phi(r), A(r) \}$ is guaranteed by
supersymmetry to be a solution of the second order field equations of
the action (\ref{GSaction}). In the case of one active scalar it is usually
straightforward to solve (\ref{gravkilling}, \ref{chikilling2}) explicitly.

We have seen that a domain wall solution of (\ref{gravkilling}),
(\ref{chikilling2}) generically has ${\cal N}=1$ 4D
Poincar\'e symmetry. Additional supersymmetries appear if the
eigenvalue $W(\phi)$ of $W_{ab}$ is degenerate, and it is known that
the Coulomb branch flow we study has maximal ${\cal N}=4$ Poincar\'e
supersymmetry. Conformal supersymmetry occurs in the boundary field
theory when the bulk geometry is exactly $AdS_5$. In this case there
are additional Killing spinors of the form \cite{Pope}
\begin{eqnarray}
\label{SCkilling}
{\epsilon}_{SC} = ( 1 - i A' x^j \gamma_j) \, e^{-A/2} \, \zeta^{(0)} \,,
\quad \quad i \gamma^r \zeta^{(0)} = - \zeta^{(0)} \,,
\end{eqnarray}
with $\zeta^{(0)}$ a constant 4D Weyl spinor.

\subsection{The GPPZ flow}

We shall primarily focus on the geometry first considered by
GPPZ \cite{GPPZ} as a
candidate dual description of a confining gauge theory.  This
background represents an operator deformation of ${\cal N}=4$ SYM, 
where the active scalar, $m$, corresponds to the addition
of equal masses for the three chiral superfields in the ${\cal N}=1$
language.  

The flow breaks the gauge symmetry from $SU(4)$ to $SO(3)$ and
the supersymmetry from 32 supercharges to 4.  As will be discussed at
length in the next section, the set of $SO(3)$-singlet fields
encompasses the gravity multiplet and two hypermultiplets, the
``active'' hyper which contains $m$, as well as the dilaton
hypermultiplet, also containing the scalar $\sigma$.\footnote{The authors
of \cite{GPPZ} additionally considered a family of flows with both $m$
and $\sigma$ active.  We restrict to the $\sigma=0$ case here, although we mention the $\sigma \neq 0$, $m=0$ background in section~\ref{CoulDiscSec}.}

The superpotential involving both scalars is
\begin{eqnarray}
\label{gppzW}
W(m,\sigma) = - \frac{3}{4} \left[\cosh \left(\frac{2m}{\sqrt{3}}
\right)+ \cosh \left( 2 \sigma \right) \right] \,,
\end{eqnarray}
and the $\sigma=0$ background that we study is described by
\begin{eqnarray}
m(r) = \frac{\sqrt{3}}{2} \log \frac{1+ e^{-r/L}}{1-e^{-r/L}} \,, \quad \quad
A(r) = \frac{1}{2} \left[ \frac{r}{L} + 
\log \left(2 \sinh \frac{r}{L} \right)\right] \,,
\end{eqnarray}
containing a singularity at finite proper distance located at $r=0$.
The variable that proves convenient for solving the various
fluctuation equations is \cite{DF}
\begin{eqnarray}
\label{udef}
u \equiv 1 - e^{-2r/L} \,,
\end{eqnarray}
in terms of which
\begin{eqnarray}
\label{symreference}
W = - \frac{3}{2u} \,, \quad
{\partial W \over \partial m} = - \sqrt{3} \; \frac{\sqrt{1-u}}{u} \,,  \quad
e^{2A} = \frac{u}{1-u} \,, \quad 
\frac{du}{dr} = \frac{2}{L} \, (1-u) \,. 
\end{eqnarray}
In the $u$-variable, the
boundary is at $u=1$ and the singularity at $u=0$.

\subsection{The Coulomb branch flow}
\label{ReviewCoulombSec}

The other background we will consider is a Coulomb branch deformation,
studied in \cite{FGPW2,BS} and called the $n=2$ flow in
\cite{FGPW2}.  It has a 10D lift to a geometry surrounding
a 2D disk of D3-branes, and preserves the symmetry $SO(4)
\times SO(2) \cong SU(2) \times SU(2) \times U(1)$.  The sixteen
supercharges dual to ordinary supersymmetries (\ref{SUSYkilling}) are
all preserved, while the superconformal supercharges (\ref{SCkilling})
are broken.

The superpotential is
\begin{eqnarray}
\label{coulW}
W(\varphi) = - e^{-2 \varphi/\sqrt{6}} - \frac{1}{2} \,
e^{4 \varphi/\sqrt{6}} \,,
\end{eqnarray}
and $\varphi \rightarrow - \infty$ as one approaches the interior.
The convenient variable for studying fluctuations in this flow is related to
the field $\varphi$ by \cite{FGPW2}
\begin{eqnarray}
v \equiv e^{\sqrt{6} \, \varphi} \,.
\end{eqnarray}
The boundary is at $v=1$, and a curvature singularity appears at
$v=0$.  The
solution for the flow is given by
\begin{eqnarray}
W = - \frac{1}{2} \,\frac{v + 2}{v^{1/3}} \,, \quad 
{\partial W \over \partial\varphi}
= \frac{2}{\sqrt{6}} \, \frac{1-v}{v^{1/3}} \,, \quad 
e^{2A} = \frac{\ell^2}{L^2} \, \frac{v^{2/3}}{1-v} \,,  \quad
\frac{dv}{dr} = \frac{2}{L} \, v^{2/3} \, (1-v) \,. 
\end{eqnarray}
The length scale $\ell$ is the radius of the disc of D3-branes.

One difference between the geometries of the two flows we study
is the behavior of radial null geodesics ($dt =-e^{-A(r)} dr$)
departing any interior point. These reach the singularity in
finite time $t$ for the GPPZ flow but take infinite time for
the Coulomb branch flow.

\section{The $SO(3)$-invariant sector of ${\cal N}=8$ gauged supergravity 
in five dimensions}
\label{InvariantSec}

It was shown in \cite{PW} that the $SO(3)$-invariant sector of the ${\cal
N}=8$ supergravity in five dimensions is described by an ${\cal N}=2$
gauged supergravity coupled to two hypermultiplets with the scalar
fields parameterizing the quaternionic manifold
\begin{equation}
\label{gtwomanfld}
{\cal Q}_0\eql {G_{2(2)}\over SO(4)}\,.
\end{equation}
In this section we present the linearized action and supersymmetry
transformation rules of this theory, which are derived by performing
an explicit truncation of the ${\cal N}=8$ supergravity to the
$SO(3)$-singlet fields.%
\footnote{The quaternionic coset $G_{2(2)}/SO(4)$ was
first studied from a different vantage point in \cite{FS} and later in \cite{BC}. In particular, the
latter reference contains an explicit realization of the $G_{2(2)}$
generators. The gauged theory which we obtain here at the linearized
level should be a particular case of gauged ${\cal N}=2$ $D=5$
supergravity coupled to matter recently constructed in
\cite{CDA}. However, we have not derived a precise mapping between the
two cases. }
Our starting point is the ${\cal N}=8$
theory \cite{PPvN,GRWlett} as formulated in \cite{GRW}, which the
reader should consult for conventions and further details (see, also,
Appendix A of \cite{FGPW1}).

We begin our discussion with a fairly detailed treatment of the scalar
sector, which requires an explicit parameterization of the scalar
manifold, ${\cal Q}_0$, and for which the linearized action is by far the
most difficult to extract. This is followed by a summary of results
for the other sectors, and finally the linearized supersymmetry
transformations.

\subsection{The scalar fields} 
\label{ScalarSubsec}

Recall that the scalar fields of the $\cN=8$ gauged supergravity in
five dimensions are given by a nonlinear $\sigma$-model on the
noncompact coset manifold $E_{6(6)}/USp(8)$. The basic object here is
the coset representative, $(\cV^{IJ\,ab},\cV_{I\alpha}{}^{ab})$,
called the ``27-bein,'' which is a matrix of $E_{6(6)}$ in a
27-dimensional real representation.  The {\bf 27} is conveniently
written in the $SL(6,R)\times SL(2,R)$ basis as $(z_{IJ},z^{I\al})$,
where $I,J = 1, \ldots, 6$ and $\alpha, \beta = 1,2$, and
$z_{IJ}=-z_{JI}$ \cite{GRW}. The corresponding infinitesimal action of
$E_{6(6)}$ is then given by (see, \cite{Crem} and, in particular,
Appendix~A of \cite{GRW})
\begin{eqnarray}
\label{esixtran}
\delta z_{IJ}&=&-\La^K{}_Iz_{KJ}-\La^K{}_Jz_{IK}+\Si_{IJK\beta}
 z^{K\beta}\,, \\
\delta z^{I\al}&=&\La^I{}_Kz^{K\al}+\La^\al{}_\beta z^{I\beta}+\Si^{KLI\al}
z_{KL}\,,  \nonumber
\end{eqnarray}
where $\La^I{}_J$, $\La^\al{}_\beta$ correspond to $SL(6,R)$ and
$SL(2,R)$ transformations respectively, and $\Si_{IJK\al} = \fracs16
\epsilon_{IJKLMN} \epsilon_{\al\beta}\Si^{LMN\beta}$.  The gauge group
$SO(6)$ is generated by $\La^I{}_J=-\La^J{}_I$.

Various $SU(2)$ invariant truncations of the theory can be obtained by
selecting a particular $SU(2)$ subgroup of the $SO(6)$ gauge group and
then restricting to the singlet fields \cite{KPW,FGPW1,PW}.  The
truncation relevant for the discussion of the GPPZ flow is obtained by
taking the obvious maximal $SO(3)$ subgroup that is diagonal in
$SO(3)\times SO(3)\subset SO(6)$, where the first $SO(3)$ acts on the
indices $I=1,2,3$ and the second one acts on $I=4,5,6$. The branching
rules that describe this embedding $SO(3)\subset SO(6) \cong SU(4)$
are $\bf 4\rightarrow 3\oplus 1$ and $\bf 6\rightarrow 3\oplus 3$.

As discussed in detail in \cite{KPW,FGPW1,PW}, the scalar manifold
of a truncated theory is given by a coset $C/K$, where $C$ is
the maximal subgroup of $E_{6(6)}$ that commutes with the invariance
subgroup and $K$ is its maximal compact subgroup. In the present case
the Lie algebra of $C$ can be constructed explicitly as
follows:

There are two obvious contributions  coming from the
$SL(2,R)\subset SL(6,R)$ transformations parametrized by
\begin{equation}
\label{KPparsltf}
(\La^I{}_J)\eql\left(\matrix{s_3\,{\bf 1} & (s_1-s_2)\,{\bf 1}\cr
(s_1+s_2)\,{\bf 1} & -s_3\,{\bf 1}\cr}\right)\,,
\end{equation}
where ${\bf 1}$ is a $3\times 3$ unit matrix, and from the 
$SL(2,R)_{\tau}$ transformations
\begin{equation}
\label{KPpartwb}
(\La^\al{}_\beta)\eql\left(\matrix {a_3 & a_1-a_2\cr
	a_1+a_2 & -a_3\cr}\right)\,,
\end{equation}
corresponding to ``the dilaton/axion'' field $\tau$ in five
dimensions.  The remaining contribution arises from the transformations
parametrized by the $\Sigma$-tensor that are invariant under
$SO(3)$. To this end define
\begin{eqnarray}
 \nonumber
X_{(1)}^{IJK} &\eql& \delta^{IJK}_{123}\,, \\
\label{KPxten}
X_{(2)}^{IJK}&\eql& \delta^{IJK}_{423}+
	\delta^{IJK}_{153}+\delta^{IJK}_{126}\,, \\ \nonumber
X_{(3)}^{IJK} &\eql& \delta^{IJK}_{156}+\delta^{IJK}_{426}+
		\delta^{IJK}_{453}\,, \\ \nonumber
X_{(4)}^{IJK} &\eql& \delta^{IJK}_{456}\,,
\end{eqnarray}
where $\delta^{IJK}_{MNP}={1\over
3!}(\delta^I_M\delta^j_N\delta^K_P+{\rm permutations})$ and set
\begin{equation}
\label{KPthesi}
\Si^{IJK\al}\eql 3!\,\sum_{i=1}^4 \tau^{i\al} \,X_{(i)}^{IJK}\,.
\end{equation}
One can check that the 14-parameter transformations (\ref{KPparsltf}),
(\ref{KPpartwb}) and (\ref{KPthesi}) indeed generate the $G_{2(2)}$
subalgebra of $E_{6(6)}$ \cite{PW}.  We will demonstrate this
explicitly by constructing a standard Cartan basis (see, {\eg},
\cite{Helg}).

Let us first choose the two simple roots, $\alpha_1$ and $\alpha_2$, of
$G_{2(2)}$ as
\begin{equation}
\label{gtworoots}
\alpha_1\eql (-\sqrt{{2\over 3}},0)\,, \qquad 
\alpha_2\eql (\sqrt{{3\over 2}},\sqrt{{1\over 2}})\,,
\end{equation}
so that
\begin{equation}
\label{rootseqs}
(\alpha_1,\alpha_1)\eql {2\over 3}\,,\qquad (\alpha_2,\alpha_2)\eql 2\,,
\qquad (\alpha_1,\alpha_2)\eql -1\,.
\end{equation}
The set of positive roots, $\Delta_+$,  consists of
\begin{equation}
\label{posrootsd}
\alpha_1\,,\quad \alpha_1+\alpha_2\,,\quad 2\alpha_1+\alpha_2\,,
\quad \alpha_2\,,\quad 3\alpha_1+\alpha_2\,,\quad 3\alpha_1+2\alpha_2\,,
\end{equation}
where the first three roots are short and the other three are long.
The set of all roots is then $\Delta=\Delta_+\cup(-\Delta_+)$.

The Cartan basis of $G_{2(2)}$ consists of two Cartan generators,
$H_1$ and $H_2$, and the root generators, $X_\alpha$,
$\alpha\in\Delta$. (For a root $\alpha=k_1\alpha_1+k_2\alpha_2$, we will
also use the notation $X_{(k_1 k_2)}\equiv X_\alpha$.) For a maximally
noncompact algebra, such as $G_{2(2)}$, the Cartan generators can be
chosen to be noncompact \cite{Helg}. It will be convenient here to
take them along the $m$ and $\sigma$ coordinate of the extended GPPZ
flow on the scalar manifold ${\cal Q}_0$.  The generators $H_1$ and $H_2$
are given in terms of those of
(\ref{KPparsltf})-(\ref{KPthesi}) in Table~1. By diagonalizing
the real matrices ${\rm ad}( H_i)$, $i=1,2$, we find the root
generators $X_\alpha$ satisfying $[H_i,X_\alpha]= \alpha^i
X_\alpha$. In Table~1 we have also given explicitly the positive root
generators, which we will need in the following. The remaining
(negative) root generators are obtained similarly with the result that
some coefficients change sign, see Table~2. As a consistency check one
may verify that the generators in Tables 1 and 2 can be block
diagonalized in agreement with the branching rule for $G_{2(2)}\subset
E_{6(6)}$, namely ${\bf 27}\rightarrow 3({\bf 7})\oplus 6 ({\bf 1})$.

\begin{table}
\begin{center}
\begin{tabular}{lcccccccc}
& $H_1$ & $H_2$ & $X_{(10)}$ & $X_{(11)}$ & $ X_{(21)}$ & $X_{(01)}$ &
	$X_{(31)}$ & $X_{(32)}$ {\vrule height0ex depth 1ex width0pt}\\
\hline\hline
$s_1 $ & $ 0 $ & $ 0 $ & $ 0 $ & $ 0 $ & $ 1\over 4\sqrt3 $ & $ 1\over 4 $ 
& $ 0 $ & $ 0 $ {\vrule height 3ex depth 0ex width0pt}\\
$s_2 $ & $ 0 $ & $ 0 $ & $ {1\over 4\sqrt3} $ & $ 0 $ & $ 0 $ & $ 0 $ 
& $ 0 $ & $ {1\over 4}$ \\
$s_3 $ & $ 0 $ & $ 0 $ & $ 0 $ & $ -{1\over 4\sqrt3} $ & $ 0 $ & $ 0 $
 & $ -{1\over 4} $ & $0 $ \\
$a_1 $ & $ 0 $ & $ 0 $ & $ 0 $ & $ 0 $ & $ \sqrt3\over4 $ & $ -{1\over 4} 
$ & $ 0 $ & $0 $ \\
$a_2 $ & $ 0 $ & $ 0 $ & $ -{\sqrt3\over 4} $ & $ 0 $ & $ 0 $ & $ 0 $ 
& $ 0 $ & $1\over 4 $ \\
$a_3 $ & $ 0 $ & $ 0 $ & $ 0 $ & $ -{\sqrt3\over 4} $ & $ 0 $ & $ 0 $ 
& $ {1\over 4} $ & $ 0 $ \\
$\tau^{11} $ & $ 0 $ & $ 0 $ & $ {1\over 8\sqrt6} $ & $ 0 $ 
& $ 1\over 8\sqrt6 $ & $ 
	{1\over 24\sqrt2}$ & $ 0	 $ 
& $ -{1\over 24\sqrt2} $ \\
$\tau^{12} $ & $ {1\over 8\sqrt3} $ & $ -{1\over 24} $ 
& $ 0 $ & $ 0 $ & $ 0 $ & $ 0 $ & $ 
	-{1\over 12\sqrt2} $ & $ 0 $ \\
$\tau^{21} $ & $ -{1\over 24\sqrt3} $ & $ -{1\over 24} $ 
& $ 0 $ & $ -{1\over 12\sqrt6}$ & $ 
		0 $ & $ 0 $ & $ 0 $ & $ 0  $ \\
$\tau^{22} $ & $ 0 $ & $ 0 $ & $ {1\over24\sqrt6} $ & $ 0 $
 & $ 1\over 24\sqrt6 $ & $ 
	-{1\over 24\sqrt2} $ & $ 0 $ & ${1\over 24\sqrt2} $ \\
$\tau^{31} $ & $ 0 $ & $ 0 $ & $ {1\over 24\sqrt6} $ 
& $ 0 $ & $ -{1\over 24\sqrt6} $ & $
	 {1\over 24\sqrt2} $ & $ 0 $ & $ {1\over 24\sqrt2} $ \\
$\tau^{32} $ & $ {1\over 24\sqrt3} $ & $ {1\over 24} $ & $ 0 $
 & $ -{1\over 12\sqrt6} $ & $ 
		0 $ & $ 0 $ & $ 0 $ & $ 0 $ \\
$\tau^{41} $ & $ -{1\over 8\sqrt3}$ & $ {1\over 24} $ & $ 0 $ 
& $ 0 $ & $ 0 $ & $ 0 $ & $ 
	-{1\over 12\sqrt2} $ & $ 0 $ \\
$\tau^{42} $ & $ 0 $ & $ 0  $ & $ {1\over 8\sqrt6} $ & $ 0 $ 
& $ -{1\over8\sqrt6} $ & $ 
	-{1\over 24\sqrt2} $ & $ 0 $ & $ -{1\over 24\sqrt2} $ {\vrule height2ex depth 2ex width0pt}
 \\
\hline
\end{tabular}
\label{tableone}
\caption{The Cartan and positive root generators of $G_{2(2)} \subset
E_{6(6)}$.}
\end{center}
\end{table}

The generators we have constructed are canonically normalized such
that ${\Tr}(H_i H_j) =2 \delta_{ij}$, ${\Tr} ( X_{\alpha}
X_{-\alpha})=2$ with all other traces being zero, with the traces
evaluated in the 7-dimensional fundamental representation of
$G_{2(2)}$. The antihermitean combinations $(X_\alpha-X_{-\alpha})$,
$\alpha\in\Delta_+$, span the compact subalgebra, while $H_1$ and
$H_2$ together with the combinations $(X_{\alpha}+X_{-\alpha})$,
$\alpha\in \Delta_+$, span the noncompact orthogonal complement.

\begin{table}
\begin{center}
\begin{tabular}{lcccccc}
&$   X_{(-10)} $&$ X_{(-1-1) } $&$ X_{(-2-1)} $&$ X_{(0-1)} $&$ X_{(-3-1)} 
$&$ X_{(-3-2)} $  {\vrule height0ex depth 1ex width0pt} \\
\hline\hline
$s_1 $&$   0 $&$ 0 $&$ 1\over 4\sqrt3 $&$ 1\over 4 $&$ 0 $&$ 0 $ {\vrule height 3ex depth 0ex width0pt}\\
$s_2 $&$   -{1\over 4\sqrt3} $&$ 0 $&$ 0 $&$ 0 $&$ 0 $&$ -{1\over 4}$ \\
$s_3 $&$   0 $&$ -{1\over 4\sqrt3} $&$ 0 $&$ 0 $&$ -{1\over 4} $&$0 $ \\
$a_1 $&$   0 $&$ 0 $&$ \sqrt3\over4 $&$ -{1\over 4} $&$ 0 $&$0 $ \\
$a_2 $&$   {\sqrt3\over 4} $&$ 0 $&$ 0 $&$ 0 $&$ 0 $&$-{1\over 4} $ \\
$a_3 $&$   0 $&$ -{\sqrt3\over 4} $&$ 0 $&$ 0 $&$ {1\over 4} $&$ 0 $ \\
$\tau^{11} $&$  {1\over 8\sqrt6} $&$ 0 $&$ -{1\over 8\sqrt6} $&$ 
	-{1\over 24\sqrt2}$&$ 0	 $&$ -{1\over 24\sqrt2} $ \\
$\tau^{12} $&$  0 $&$ 0 $&$ 0 $&$ 0 $&$ 
	{1\over 12\sqrt2} $&$ 0 $ \\
$\tau^{21} $&$  0 $&$ {1\over 12\sqrt6}$&$ 
		0 $&$ 0 $&$ 0 $&$ 0  $ \\
$\tau^{22} $&$  {1\over24\sqrt6} $&$ 0 $&$ -{1\over 24\sqrt6} $&$ 
	{1\over 24\sqrt2} $&$ 0 $&${1\over 24\sqrt2} $ \\
$\tau^{31} $&$  {1\over 24\sqrt6} $&$ 0 $&$ {1\over 24\sqrt6} $&$
	 -{1\over 24\sqrt2} $&$ 0 $&$ {1\over 24\sqrt2} $ \\
$\tau^{32} $&$  0 $&$ {1\over 12\sqrt6} $&$ 
		0 $&$ 0 $&$ 0 $&$ 0 $ \\
$\tau^{41} $&$  0 $&$ 0 $&$ 0 $&$ 0 $&$ 
	{1\over 12\sqrt2} $&$ 0 $ \\
$\tau^{42} $&$  {1\over 8\sqrt6} $&$ 0 $&$ {1\over8\sqrt6} $&$ 
	{1\over 24\sqrt2} $&$ 0 $&$ -{1\over 24\sqrt2} $ {\vrule height2ex depth 2ex width0pt} \\
\hline
\end{tabular}
\label{tabletwo}
\caption{The negative root generators of $G_{2(2)} \subset E_{6(6)}$.}
\end{center}
\end{table}

The usefulness of the above construction for our purposes lies in the
fundamental theorem on the Iwasawa decomposition of the maximally
noncompact Lie groups (see, {\eg}, \cite{Helg}), which implies that the
scalar coset (\ref{gtwomanfld}) can be analytically parametrized by   real
coordinates $s_i$, $i=1,2$, and $t_\alpha$, $\alpha\in\Delta_+$,
where an explicit mapping is given by the group elements
\begin{equation}
\label{iwasawa}
g(s_i,t_\alpha)\eql \exp\left(\sum_{i=1}^2 s_iH_i\right)\,
\exp(\sum_{\alpha\in\Delta_+}t_\alpha X_\alpha)\,,
\qquad s_i\,,t_\alpha\in {\bf R}\,.
\end{equation}
Given an explicit representation of the generators, the 
group element (\ref{iwasawa}) is easy to compute -- the factor on the left
is a product of commuting $O(1,1)$ ``rotations'', while the factor on the
right is an element of a solvable group%
\footnote{This is why the parameterization has been called a ``solvable 
parameterization'' in \cite{ADAFFT}.}
and thus a matrix of finite degree polynomials in the $t_\alpha$'s. 

In the following we will use instead of $s_1$ and $s_2$ the
corresponding fields of the extended GPPZ-flow \cite{GPPZ}, which are
simply given by
\begin{equation}
\label{gppztophi}
m\eql {1\over \sqrt2}\,s_1\,,\qquad \sigma\eql {1\over
\sqrt2}\,s_2\,.
\end{equation}
We will also retain in all the formulae of this section the complete
dependence on the two fields $m$ and $\sigma$.

Once the proper parameterization of the scalar coset ${\cal Q}_0$ has been
introduced, the computation of the kinetic action and the potential is
a matter of a straightforward algebra that can be carried out on a
computer. First one evaluates the group elements (\ref{iwasawa}) in a
convenient 7-dimensional representation and derives the kinetic action
using the standard $\sigma$-model techniques. Then by a basis change
one obtains the 27-bein fields $(\cV^{IJ\,ab},\cV_{I\alpha}{}^{ab} )$,
which can be substituted directly into the formulae in \cite{GRW}.
Since several calculations of this type have already been discussed in
the literature (see, {\eg} \cite{FGPW1}) we will skip the details
here.

In the absence of vector fields the kinetic action of the scalar
fields is simply given by
\begin{equation}
\label{scalarkin}
{\cal L}^0_{K}(m,\sigma;t_\alpha)
\eql {1\over 2} \left[(\partial m)^2+(\partial\sigma)^2 \right]+
{1\over 2}\sum_{\alpha\in\Delta_+} e^{2\alpha(m,\sigma)}
(\partial t_\alpha)^2\,,
\end{equation}
where $\alpha(m,\sigma)=\sqrt2 (\alpha^1m+\alpha^2\sigma)$ for a root
$\alpha$.

Using the solvable parameterization we have obtained a complete
expression for the potential. The $t_\alpha$-independent part of
the potential reproduces the result in \cite{GPPZ,DZ,PW}:
\begin{equation}
\label{zeropot}
{\cal P}^{(0)}(m,\sigma)\eql -{3\over 16}g^2\,
\left(2-{1\over 4} \cosh(4\sigma)+{1\over 4}\cosh({2m\over\sqrt3})+
	\cosh({2m\over\sqrt3}+2\sigma)+
	\cosh({2m\over\sqrt3}-2\sigma)\right)\,.
\end{equation}
The complete expression for the potential is too long to be reproduced
here.\footnote{An interested reader may enjoy it at
http://citusc.usc.edu/$\sim$pilch/Papers/g2potexp.out.} It is a sixth order
polynomial in the $t_\alpha$'s with no first order terms.  Since all
we need here is the quadratic expansion in the transverse fields
$t_\alpha$ about an arbitrary $m$ and $\sigma$ configuration, we now
restrict to this order. Define
\def\Es#1#2{{E(#1,#2)}}
\def\t_#1{t_{(#1)}}
\begin{equation}
\label{defofexp}
\Es {k_1}{k_2}\eql e^{\alpha(m,\sigma)}\qquad {\rm where} \qquad
\alpha=k_1\alpha_1+k_2\alpha_2\,.
\end{equation}
Then the quadratic term in the expansion of the potential is
\begin{equation}
\label{quadpot}
\begin{array}{rcl}
{\cal P}^{(2)}(m,\sigma;t_\alpha) &=&
-{g^2\over 16}(1+2\Es{-2}{-2}+\Es 20+2\Es 42)\,\t_{10}^2 \\
&& -{g^2 \over 32}(8-3 \Es{-4}{-2}+6\Es{-2}0-\Es 02-8\Es 22+2\Es 64)\, 
	\t_{11}^2 \\
&&-{g^2 \over 32}(8-3 \Es{-2}{-2}+6\Es{2}0-8\Es 42-\Es 62+2\Es 64)\, 
	\t_{21}^2 \\
&&+{3 g^2 \over 32}(\Es{-6}{-2}-4\Es{-2}0-\Es22)\,\t_{01}^2 \\
&&+{3 g^2 \over 32}(\Es0{-2}-4\Es20-\Es42)\,\t_{31}^2 \\
&&+{3 g^2 \over 16}(1-2\Es22-2\Es42+\Es64)\,\t_{32}^2 \\
&&+{\sqrt3 g^2 \over 16}(\Es{-2}0-6\Es22+\Es64)\t_{21}\t_{01} \\
&&+{\sqrt3 g^2 \over 16}(\Es20-6\Es42+\Es64)\,\t_{11}\t_{31}\,.
\end{array}
\end{equation}
Note that up to this order the scalar action is diagonal in $\t_{10}$
and $\t_{32}$ and there are two $2\times 2$ blocks in
$\t_{11}/\t_{31}$ and $\t_{21}/\t_{01}$, respectively.

The potential in (\ref{zeropot}) and (\ref{quadpot}) involves all
eight $SO(3)$-singlet fields, but is still invariant under the
$SL(2,R)_\tau$ and an $O(2)$ subgroup of the gauge group. The former
symmetry is given by the transformations (\ref{KPpartwb}) acting on the
$SL(2,R)$ index $\alpha$ of the 27-bein. This is a manifest local
symmetry of the potential of the full ${\cal N}=8$ theory
\cite{GRW}. However, its action on the $SO(3)$-singlet fields in our
solvable parameterization is quite complicated as might be inferred
from the explicit form of the $SL(2,R)_\tau$ generators, given by
\begin{equation}
\label{sl2tau}
\begin{array}{rcl}
T_+ &=& 
-{\sqrt 3\over
2}(X_{(11)}+X_{(-1-1)})+{1\over 2}(X_{(31)}+X_{(-3-1)})\,, \\
T_- &=& {\sqrt3\over 2}(X_{(21)}+X_{(-2-1)})-{1\over
2}(X_{(01)}+X_{(0-1)})\,, \\
T_3 &=& -{\sqrt{3}\over 2}(X_{(10)}-X_{(-10)})+{1\over
2}(X_{(32)}-X_{(-3-2)})\,, 
\end{array}
\end{equation}
which act on the scalar coset, ${\cal Q}_0$, by right
multiplication of (\ref{iwasawa}). The generator of the $O(2)$ symmetry
is given in (\ref{uonegen}) below. 

If one is interested in studying the potential alone, it is more
advantageous to use another parameterization of ${\cal Q}_0$ and gauge
fix all symmetries. This was done in \cite{PW}, where the potential
for the $SO(3)$-singlet fields was shown to depend on $m$ and $\sigma$
and two other fields, which together parametrized a quotient of
$SL(3,R)/SU(2)$ by an $O(2)$ action. However, one must remember that
the above local symmetries of the potential are {\it not} local
symmetries of the full scalar action and thus a careful treatment of
the dynamics of the $SO(3)$-singlet sector requires that we consider
all eight fields. For this purpose the solvable parameterization employed
here is quite convenient.

An action for some of the $SO(3)$-invariant scalars of the gauged
${\cal N}=8$ theory was also derived and studied in \cite{DZ}.  The
authors restrict to a 6-dimensional scalar subspace, and their scalar
potential depends only on 2 of the 6 scalars, not including the
dilaton nor any field mixing with it.  A direct comparison of our
results with \cite{DZ} is rather difficult because complicated field
redefinitions are required (although we have reproduced the truncated
action in their parameterization). Nevertheless we believe that their
results are incorrect because there is no natural 6-dimensional
subspace of the coset $G_{2(2)}/SO(4)$ nor any symmetry to enforce
such a truncation. A more precise criticism is that their kinetic
action for the dilaton gives a fluctuation equation (about the GPPZ
flow) which is not hypergeometric, as opposed to those of all other
$SO(3)$-singlet fluctuations we study. As shown in
section~\ref{ScalarSec} below, the dilaton and axion emerge precisely
from the mixed $2\times2$ sectors of the quadratic potential
(\ref{quadpot}).  This is a consequence of the action and equations of
motion for these fields, and doesn't depend on the form of the
background $\{ m(r), \sigma(r) \}$ or the presence of
supersymmetry. For these reasons the 6-dimensional truncation in
\cite{DZ} appears to be inconsistent and the application to chiral
symmetry breaking should be reexamined.

\subsection{The vector field and the vector/scalar coupling}
\label{vectorscalarcplng}

The commutant of $SO(3)$ in the gauge group $SO(6)$ consists of a
single $U(1)$ generated by the transformations with
$\La^1{}_4=\La^2{}_5=\La^3{}_6$ and the corresponding generator in
the Cartan basis given by
\begin{equation}
\label{uonegen} 
T_R\eql -{{\sqrt 3\over
2}}(X_{(10)}-X_{(-10)})-{3\over 2}(X_{(32)}-X_{(-3-2)})\,.
\end{equation}
We will denote this subgroup by $U(1)_R$ to emphasize that it
corresponds to the truncation of the $R$-symmetry group on the field
theory side. This symmetry is, however, broken by the background
fields $m$ and $\sigma$ and as a result the corresponding vector
field, $B_\mu$, develops an $m$- and $\sigma$-dependent mass-term. A
tedious expansion of the $\cN=8$ $D=5$ action to the quadratic order
in $B_\mu$ yields
\begin{equation}
\label{vectact}
{\cal L}(B_\mu;m,\sigma)\eql 
-{3\over 4} F_{\mu\nu}F^{\mu\nu} +{3\over 16} g^2
 \left(3\cosh(4\sigma)+\cosh({4m\over \sqrt{3}})-4\right)\,B_\mu B^\mu\,.
\end{equation}
In deriving (\ref{vectact}) we have used that there are no
$SO(3)$-singlet antisymmetric tensor fields that might mix with
$B_\mu$. One can use (\ref{gppzW}) and the flow equations
(\ref{gravkilling}), (\ref{chikilling2}) to show that the vector mass
is $m_B^2 = -2 A''(r)$.  We will show in the next section that this
result is quite generic. 

The coupling between the vector and the scalar fields up to the
quadratic order in the fluctuations is given by
\begin{eqnarray}
\label{vectsccop}
&&{\cal L}^{(2)}(B_\mu,t_\alpha ; m,\sigma) = \\
\qquad
&&B_\mu\left(3 (\partial^\mu \sigma) \t_{32}-(\partial^\mu m) \t_{10}
+{\sqrt 3\over 4}(1-e^{-4m/\sqrt3})\partial^\mu \t_{10}
+{3\over 4}(1-e^{4\sigma})\partial^\mu \t_{32}\right)
\nonumber \,,
\end{eqnarray}
which is in agreement with the usual gauging of a nonlinear
$\sigma$-model as we show in more detail in section~\ref{ScalarSec}.

\subsection{The fermion fields}
\label{fermionfields}

The $SO(3)$ truncation of the fermion sector yields one symplectic
Majorana pair of spin-3/2 fields, $\tilde{\psi}_\mu^a$, $a=1,2$,
and two pairs of spin-1/2 fields, $\tilde{\chi}^a$,
$a=1,\ldots,4$. This follows from the counting of $SO(3)$ singlets in
the branching of ${\bf 8}$ and ${\bf 48}$ of $USp(8)$. We use
formulae in Appendix A of \cite{GRW} for the embedding of $SO(3)$ into
$USp(8)$ (with the $SO(7)$ gamma matrices given in Appendix C of
\cite{FGPW1}) to determine those $SO(3)$-singlet fields in terms of
the original fields of the ${\cal N}=8$ theory:
\begin{equation}
\label{thepsising}
\tilde{\psi}_\mu^1\eql \psi_\mu^3 \,,\qquad
\tilde{\psi}_\mu^2\eql \psi_\mu^7 \,,
\end{equation}
which mirrors the truncation of the supersymmetry parameter and 
\begin{equation}
\label{spinhalfsing}
\begin{array}{rcl}
\tilde{\chi}^1 &=& \chi^{124} \,,\\
\tilde{\chi}^2 &=&\chi^{128}  \eql -\chi^{146}\eql \chi^{245}\,,\\
\tilde{\chi}^3 &=& \chi^{168}  \eql -\chi^{258}\eql \chi^{456} \,,\\
\tilde{\chi}^4 &=& \chi^{568}  \,.
\end{array}
\end{equation}
In (\ref{spinhalfsing}) we have used that $\chi^{abc}\eql \chi^{[abc]|}$
are antisymmetric and symplectic traceless $USp(8)$ tensors and listed
the independent components only. 

The reduction of the fermion action is now rather easy and we find
that further redefinitions of the spin-1/2 fields are required
to bring it into canonical form. Let us define 
\def\xxi{\xi}
\begin{eqnarray}
\label{canfermions}
\xxi^1&=& {\sqrt 3\over 2}(\tilde\chi{}^2+\tilde\chi{}^4)\,,\qquad
\xxi^2\eql {\sqrt 3\over 2}(\tilde\chi{}^1+\tilde\chi{}^3)\,,\\
\nonumber
\xxi^3&=& -{1\over 2}(3\tilde\chi{}^2-\tilde\chi{}^4)\,,\qquad
\xxi^4\eql {1\over 2}(\tilde\chi{}^1-3\tilde\chi{}^3)\,.
\end{eqnarray}
In terms of those fields the fermion kinetic action becomes
\begin{eqnarray}
\label{feractke}
{\cal L}_K^{(2)}(\psi_\mu^a,\xi^a) &=&
-\frac{i}{2} \left( \bar\psi{}^1_\mu \gamma^{\mu\nu\rho}D_\nu\psi_\rho^2-
\bar\psi{}^2_\mu \gamma^{\mu\nu\rho}D_\nu\psi_\rho^1 \right)
\\
&&-\frac{i}{2} \left(
\bar\xi^1 \gamma^\mu D_\mu \xi^2 -
\bar\xi^2 \gamma^\mu D_\mu \xi^1 +
\bar\xi^3 \gamma^\mu D_\mu \xi^4 -
\bar\xi^4 \gamma^\mu D_\mu \xi^3 \right) \,,
\nonumber
\end{eqnarray}
where we dropped the tilde over the spin-3/2 field. The
truncation of the mass terms is more complicated as it requires
evaluating the $USp(8)$-tensors of the ${\cal N}=8$ theory. We find
\begin{eqnarray}
\label{themassfer}
\nonumber
{\cal L}_M^{(2)}(\psi_\mu^a,\xi^a)&=&
-{i g\over 4}  W(m,\sigma) (\bar\psi_\mu^1\gamma^{\mu\nu}\psi_\nu^1+
\bar\psi_\mu^2\gamma^{\mu\nu}\psi_\nu^2)
\\
&& 
-{i g\over 16}  \left(\cosh({2m\over\sqrt3})-3\cosh(2\sigma)\right)
\left(  \bar\xi^1 \xi^1 +
  \bar\xi^2 \xi^2
-3 \bar\xi^3 \xi^3 
-3 \bar\xi^4 \xi^4 
 \right)
\\ \nonumber 
&&
+{1\over\sqrt2}m'(r) \left(
\bar\psi^1_\mu \gamma^r\gamma^\mu\xxi^1+
\bar\psi^2_\mu \gamma^r\gamma^\mu\xxi^2
\right) 
+{1\over \sqrt{2}} {\sigma}'(r) \left( 
\bar\psi^1_\mu \gamma^r\gamma^\mu\xxi^3+\bar\psi_\mu^2
\gamma^r\gamma^\mu\xxi^4\right) \\ \nonumber
&& - {\sqrt3 g\over 4\sqrt2}  \sinh({2m\over\sqrt3})\left(
\bar\xxi^2\gamma^\mu\psi_\mu^1-
\bar\xxi^1\gamma^\mu\psi^2_\mu\right) 
-{\sqrt3 g\over 4\sqrt2} \sinh(2\sigma) \left(
\bar\xxi^4\gamma^\mu\psi_\mu^1-
\bar\xxi^3\gamma^\mu\psi^2_\mu\right) \,,
\end{eqnarray}
where $W(m,\sigma)$ is the superpotential (\ref{gppzW}).
This may be further simplified using the flow equations
(\ref{gravkilling}), (\ref{chikilling2}), which we will do in
section~\ref{FermionSec}.

\subsection{The supersymmetry transformations}
\label{susytrans}

We will now verify that to the linear order we have indeed obtained a
gauged ${\cal N}=2$ supergravity coupled to two hypermultiplets. For
the supergravity multiplet we
find:
\def\scp#1#2{{E(#1,#2)\partial_\mu t_{(#1#2)}}}
\begin{equation}
\label{sugramul}
\begin{array}{rcl}
\delta e_\mu^m &=&-i(\bar \epsilon^1\gamma^m \psi_\mu^2-
	\bar \epsilon^2\gamma^m \psi_\mu^1)\,,  \\
\delta\psi_\mu^1 &=& D_\mu\epsilon^1
  -{1\over 6} gW(m,\sigma)\gamma_\mu\epsilon^1
  -g B_\mu W(m,\sigma) \epsilon^2+
{1\over 4}F_{\nu\rho}(\gamma^{\nu\rho}\gamma_\mu+
2\gamma^\nu\delta_\mu^\rho )\epsilon^1 \\
&&+{ig\over 2}\left(\scp01-\sqrt3\scp21\right)\epsilon^1\\
&&-{g\over 2}\left( \sqrt3\scp10+\sqrt3 i \scp11-i\scp31+\scp32
 \right)\epsilon^2\,,\\
\delta\psi_\mu^2 &=& D_\mu\epsilon^2+
{1\over 6}gW(m,\sigma)\gamma_\mu\epsilon^2 +g B_\mu W(m,\sigma)\epsilon^1+
{1\over 4}F_{\nu\rho}(\gamma^{\nu\rho}\gamma_\mu+
2\gamma^\nu\delta_\mu^\rho )\epsilon^2 \\
&&+{g\over 2}\left ( \sqrt3 \scp10-\sqrt3 i\scp11+i\scp31+\scp32
\right)\epsilon^1\\
&&-{ig\over 2}\left(\scp01-\sqrt3\scp21\right)\epsilon^2\,,\\
\delta B_\mu &=& - {i\over 4} \,(\bar\epsilon{}^1\psi_\mu^2-
\bar\epsilon{}^2\psi_\mu^1)\,. 
\end{array}
\end{equation}
To present the supersymmetry transformations for the
hypermultiplets, let us define rescaled variation of $t_\alpha$, cf.\
(\ref{scalarkin}),
\begin{equation}
\label{rescalet}
\tilde{\delta}t_\alpha\eql e^{\alpha(m,\sigma)}\delta t_\alpha\,,
\end{equation}
in terms of which the result has a compact form: 
\begin{equation}
\label{susyone}
\begin{array}{rcl}
\delta m &=& - {1\over\sqrt2} (\bar\epsilon^1 \xi^1 +
\bar\epsilon^2 \xi^2 ) \,, \\
\tilde{\delta}t_{(10)} &=& - \bar\epsilon^1 \xi^2 + 
	\bar\epsilon^2 \xi^1 \,, \\
{\sqrt 3\over 2}\tilde{\delta}t_{(01)}+{1\over 2}\tilde{\delta}t_{(21)}
&=& i(\bar\epsilon^1 \xi^1 - \bar\epsilon^2 \xi^2) \,, \\
{\sqrt 3\over 2}\tilde{\delta}t_{(31)}+{1\over 2}\tilde{\delta}t_{(11)}
&=&
-i(\bar\epsilon^1 \xi^2+ \bar\epsilon^2 \xi^1) \,, 
\end{array}
\end{equation}
and
\begin{equation}
\label{susytwo}
\begin{array}{rcl}
\delta \sigma &=& -{1\over\sqrt2}
  (\bar\epsilon^1 \xi^3 +\bar\epsilon^2 \xi^4 ) \,, \\
\tilde{\delta}t_{(32)} &=& (\bar\epsilon^1 \xi^4 - \bar\epsilon^2 \xi^3) 
\,,\\
{1\over 2}\tilde{\delta}t_{(01)}-{\sqrt3\over 2}\tilde{\delta}t_{(21)}
&=& i(\bar\epsilon^1 \xi^3 - \bar\epsilon^2 \xi^4) \,,\\
-{1\over 2}\tilde{\delta}\t_{31}+{\sqrt{3}\over 2}\tilde{\delta}t_{(11)}&=&
-i(\bar\epsilon^1 \xi^4+ \bar\epsilon^2 \xi^3) \,.
\end{array}
\end{equation}

The particular combinations of the scalar fields we have displayed
here turn out to be crucial for the diagonalization of the field
equations in section~\ref{ScalarSec}.

Similarly, for the variation of the spin-1/2 fields let us use
the notation
\begin{equation}
\label{redeft}
\tilde{\dslash} t_\alpha\eql  e^{\alpha(m,\sigma)}\dslash t_\alpha\,.
\end{equation}
Then
\def\tt#1#2{t_{(#1#2)}}
\begin{equation}
\label{susyonef}
\begin{array}{rcl}
\delta\xi^1 &=& {i\over \sqrt2}\dslash m \,\epsilon^2 
-{1\over 4}(\sqrt3\tilde{\dslash}\tt01 +\tilde{\dslash}\tt21)\,\epsilon^2
+{i\over 2}\tilde{\dslash}\tt10\,\epsilon^1
-{1\over 4}(\sqrt3\tilde{\dslash}\tt31+\tilde{\dslash}\tt11)\,\epsilon^1
\\
  &&- {ig\sqrt3\over 4\sqrt2}\,\sinh({2m\over\sqrt3})\,\epsilon^1
     +{ig\sqrt3\over 2\sqrt2}\, \Bslash \,\sinh({2m\over\sqrt3})
	\,\epsilon^1
\\
  && +
{g\sqrt3\over 16} (3 E(-3, -1) + E(1, 1))\,\tt01\, \epsilon^1   -
{g\over 16} (3 E(-2, -1) + 3 E(0, 1) - 2 E(2, 1))\,\tt11\, \epsilon^2 
\\  && + 
{g\over 16}(3 E(-1, -1) - 2 E(1, 1) + 3 E(3, 1))\,\tt21\, \epsilon^1  - 
{g\sqrt3\over 16} (3 E(0, -1) + E(2, 1))\, \tt31\,\epsilon^2 
\\ && + {ig\over 4} \,\tt10\, \epsilon^2 
\end{array}
\end{equation}
\begin{equation}
\label{susytwof}
\begin{array}{rcl}
\delta\xi^2 &=& -{i\over \sqrt2}\dslash m \,\epsilon^1 
-{1\over 4}(\sqrt3\tilde{\dslash}\tt01+\tilde{\dslash}\tt21)\,\epsilon^1
+{i\over 2}\tilde{\dslash}\tt10\,\epsilon^2
+{1\over 4}(\sqrt3\tilde{\dslash}\tt31+\tilde{\dslash}\tt11)\,\epsilon^2
\\ &&- {ig\sqrt3\over 4\sqrt2}\,\sinh({2m\over\sqrt3})\,\epsilon^2
    +{ig\sqrt3\over 2\sqrt2}\, \Bslash \,\sinh({2m\over\sqrt3})
	\,\epsilon^2
\\ && -
  {g\sqrt3\over 16} (3 E(-3, -1) + E(1, 1))\, \tt01\, \epsilon^2  - 
 {g\over 16} (3 E(-2, -1) + 3 E(0, 1) - 2 E(2, 1))\,\tt11\, \epsilon^1  
\\ &&-
 {g\over 16} (3 E(-1, -1) - 2 E(1, 1) + 3 E(3, 1))\,\tt21\, \epsilon^2  - 
  {g\sqrt3\over 16} (3 E(0, -1) + E(2, 1))\,\tt31\, \epsilon^1 
\\ &&-
{ig\over 4}\, \tt10\, \epsilon^1 \,.
\end{array}
\end{equation}
and
\begin{equation}
\label{susythree}
\begin{array}{rcl}
\delta\xi^3 &=& {i\over \sqrt2}\dslash\sigma\,\epsilon^2 
-{1\over 4}(\tilde{\dslash}\tt01-\sqrt3\tilde{\dslash}\tt21)\,\epsilon^2
-{i\over 2}\tilde{\dslash}\tt32\,\epsilon^1
+{1\over 4}(\tilde{\dslash}\tt31-\sqrt3\tilde{\dslash}\tt11)\,\epsilon^2
\\ && - {3ig\over 4\sqrt2}\,\sinh(2\sigma)\,\epsilon^1
    +{3ig\over 2\sqrt2}\, \Bslash \,\sinh(2\sigma)\,\epsilon^1
\\ && -
{3 g\over 16}(E(-3, -1) - E(1, 1))\,\tt01\, \epsilon^1  - 
  {g\sqrt3\over 16} (-3 E(-2, -1) + E(0, 1) + 2 E(2, 1))\,\tt11\, \epsilon^2  
\\ && -
  {g\sqrt3\over 16} (-3 E(-1, -1) + 2 E(1, 1) + E(3, 1))\, \tt21\,\epsilon^1 
 -{3 g\over 16} (E(0, -1) - E(2, 1)) \,\tt31\,\epsilon^2   
\\ && - 
  {3 ig\over 4}\,\tt32\,\epsilon^2 \,,
\end{array}
\end{equation}
\begin{equation}
\label{susyfour}
\begin{array}{rcl}
\delta\xi^4 &=& -{i\over \sqrt2} \dslash\sigma\,\epsilon^1 
-{1\over 4}(\tilde{\dslash}\tt01-\sqrt3\tilde{\dslash}\tt21)\,\epsilon^1
-{i\over 2}\tilde{\dslash}\tt32\,\epsilon^2
-{1\over 4}(\tilde{\dslash}\tt31-\sqrt3\tilde{\dslash}\tt11)\,\epsilon^2
\\
  &&  - {3ig\over 4\sqrt2}\,\sinh(2\sigma)\,\epsilon^2
    +{3ig\over 2\sqrt2}\, \Bslash \,\sinh(2\sigma)\,\epsilon^2
\\ &&
+{3 g\over 16} \left(E(-3, -1) - E(1, 1)\right)\, \tt01 \,\epsilon^2 
-  {g\sqrt3\over 16}   (-3 E(-2, -1) + E(0, 1) + 
2 E(2, 1)) \,\tt11\,\epsilon^1 
\\ &&
- {g\sqrt3 \over 16} (3 E(-1, -1) - 2 E(1, 1) - E(3, 1)) \,\tt21 \,\epsilon^2
 -  {3 g\over 16} (E(0, -1) - E(2, 1))\,\tt31\, \epsilon^1 
\\ &&
 + {3 ig\over 4}\, \tt32\,  \epsilon^1 \,,
\end{array}
\end{equation}
When the background satisfies the flow equations (\ref{gravkilling}),
(\ref{chikilling2}) the zeroth-order terms in $\delta \psi_\mu$ and
$\delta \xi$ will vanish for $\epsilon$ a Killing spinor,
as given by (\ref{SUSYkilling}).

\section{Physical modes of the gravity multiplet}
\label{TTSec}

In this section we consider the physical (\ie transverse, traceless)
modes of the metric fluctuation $h_{\mu \nu} \equiv e^{-2A} \, \hat\delta
g_{\mu \nu}$, gravitino $\psi^a_\mu$, and $U(1)_R$ photon $B_\mu$,
which are dual to the superconformal current multiplet containing the
stress tensor, supercurrent and $R$-current.  These TT modes are
non-vanishing only for transverse values of the indices, \ie $\mu,\nu
\rightarrow i,j$, and they will decouple from other modes and from
lower spin fluctuations; we denote them $\{ \hat{h}_{ij}, \hat\psi_i,
\hat{B}_i \}$.

In the GPPZ background, these modes obey the linear equations of motion: \ba
\label{ttgrav} R^{(1)}_{ij}(\hat{h}) + \fracs{4}{3} V(m(r)) \, \hat{h}_{ij}
&=& 0 \,, \\
\label{ttpsi} \gamma^{\mu\nu\rho}\hat{D}_\nu \hat\psi_\rho &=& 0 \,, \\
\label{tta} D^\mu \hat{F}_{\mu\nu}-2A''(r) \hat{B}_\nu &=& 0 \,,
\ea
where we combined the symplectic Majorana gravitini into a complex field
$\psi_\rho \equiv \psi_\rho^1 + i \psi_\rho^2$, and
\be
\label{dhat}
\hat{D}_\nu \equiv D_\nu - \frac{ig}{6} W \gamma_\nu \,.
\ee
We will not write the linearized Ricci operator $R^{(1)}_{ij}$ in
detail (see for example \cite{DFGK}), since it is by now well-known
(as derived in \cite{BS,DFGK} for a general domain wall and earlier in a
special case \cite{RS}) that the TT modes of $\hat{h}_{ij}$ obey the same
equation as that of a massless scalar,
\be\label{phieq}
D^\mu \partial_\mu {f}(x^i,r) = \left[- \partial_r^2 - 4A' 
\partial_r  + e^{-2A} \square \right] {f}(x^i,r) = 0 \,, \qquad \hat{h}_{ij}
\equiv v_i v_j {f}(x^i,r) \,,
\ee
with $v_i$ independent of $r$ and transverse.  The TT constraints then
require that the $v_i$ are constructed from three independent polarization
vectors (as described in detail in (21) of \cite{DF}). In this way one
can build the five independent physical modes of $\hat{h}_{ij}$.  The solution
for the GPPZ flow was found to be \cite{AGPZ,DF}:
\begin{eqnarray}
{f} = (1-u)^2 \, F \left( 2 + \frac{pL}{2}, 2 - \frac{pL}{2} ;2;u\right) \,,
\end{eqnarray}
where the radial variable $u$ is defined in (\ref{udef}) and
$F(a,b,c;z)$
denotes a standard hypergeometric function \cite{Bate}.  The associated
spectrum contains discrete states with momenta $p^2 L^2 = 4 (n+2)^2$,
in agreement with (\ref{Jspec}).  It should be noted the
$SO(3)$-singlet supergravity theory does not contain a massless scalar
field; ${f}$ is just a convenient auxiliary quantity.

Since $\hat\psi_\mu$ and $\hat{B}_\mu$ are SUSY partners of
$\hat{h}_{ij}$, we expect that they can also be expressed in terms of
the massless scalar ${f}$, and it is our purpose to show how to do
this.  We use supersymmetry transformation rules to relate solutions
of the equations of motion (\ref{ttgrav}) -(\ref{tta}).  The necessary
parts of the transformations (\ref{sugramul}), rewritten in terms of
complex spinors, are \ba
\label{trfe} \delta e_\mu^{\hat{m}} &=& \bar{\epsilon} \gamma^{\hat{m}} \psi_\mu \,, \\
\label{trfpsi} \delta \psi_\mu &=& \hat{D}_\mu \epsilon \,, \\
\label{trfa} \delta B_\mu &=& {1\over 4} \bar{\epsilon} \, \psi_\mu \,,
\ea
where $\epsilon$ is the Killing spinor of (\ref{SUSYkilling}). 

Let $\hat{h}_{ij}(x^i,r)$ be any solution of (\ref{ttgrav}).  The
corresponding fluctuation of the frame can be chosen as
\be\label{flucte} \hat\delta \hat{e}^{\hat{k}}_j= \fracs12 e^{A(r)} 
\eta^{{\hat{k}}i}
\hat{h}_{ij}(x^k,r) \,, \ee where all other components vanish and
$\eta^{{\hat{k}}i}$ is the $4\times4$ Minkowski metric. Let $\hat\delta
\omega_{\mu{\hat{k}}{\hat{l}}}$ 
denote the corresponding fluctuation of the spin
connection. It is then guaranteed that \be \hat{\psi}_\mu =
(\partial_\mu + \frac{1}{4} \,\hat\delta \omega_{\mu {\hat{k}}{\hat{l}}} 
\gamma^{{\hat{k}}{\hat{l}}} +
\frac{i}{2} A'(r) \gamma_{\hat{k}} \hat\delta e^{\hat{k}}_\mu) \, \epsilon \,,
\ee 
is a
solution of (\ref{ttpsi}). It is now straightforward to use the
specific form of $\hat\delta \omega_{\mu{\hat{k}}{\hat{l}}}$ 
and show that the radial
component $\hat{\psi}_5$ vanishes, and that all terms in which
$\hat{h}_{ij}$ is not differentiated cancel in $\hat{\psi}_i$. One can
then use the transverse traceless property of $\hat{h}_{ij}$ to bring
$\hat{\psi}_i$ to the form \be\label{psih} \hat{\psi}_i = \frac{1}{4}
\, e^{2A(r)} \gamma^j (\gamma^\mu\partial_\mu \hat{h}_{ij}) \epsilon
\ee Despite appearances this form is covariant since the factor
$e^{2A(r)}$ can be moved to the right of the derivative. One then
finds that $\partial_\mu$ is replaced by the covariant $D_\mu$ acting
on the tensor fluctuation $\hat\delta g_{ij} = e^{2A(r)} \hat{h}_{ij}$.
It is easy to see that $\gamma^i \hat{\psi}_i$ and $g^{ij} D_i
\hat{\psi}_j$ vanish, so these gravitino modes are transverse
traceless.

The final step is to express $\hat{\psi}_i$ in terms of the scalar
${f}$. To do this we simply substitute $\hat{h}_{ij} = v_i v_j {f}$
in (\ref{psih}) to obtain
\be\label{psiphi}
\hat{\psi}_i = -\frac{1}{4} \, e^{2A(r)} v_i 
(\gamma^\mu\partial_\mu {f}) (\gamma^j v_j) 
  \epsilon \,.
\ee
It is convenient to define a new spinor $\tilde{\epsilon} = (\gamma^j
v_j) \, \epsilon$. This might be called an anti-Killing spinor,
since the form of $\tilde{\epsilon}$ is the same as (\ref{SUSYkilling}) but
with opposite chirality of the constant
spinor $\eta^{(0)}$, and consequently $(D_\mu + \frac{ig}{6} W \gamma_\mu) \,
\tilde{\epsilon} = 0$.  Dropping an irrelevant constant we can then write
\be\label{psiphi1}
\hat{\psi}_i = e^{2A(r)} v_i \, (\gamma^\mu\partial_\mu {f}) \,
\tilde{\epsilon} \,.
\ee
We have verified that this form is indeed a solution of (\ref{ttpsi}).
The next issue is to count the linearly independent modes of the form
(\ref{psiphi1}). This requires detailed analysis in which specific
polarization vectors are paired with choices of $\eta$ to create modes
of overall half-integer helicity.  There are a total of 4 independent
modes, as expected.\footnote{It may appear that the form (\ref{psiphi})
contains more modes. However, one must take into account the fact that
$\gamma^j v_j$ has zero modes when $v_j$ is a circular polarization
vector.}

We proceed now to treat the graviphoton field $B_\mu$
in the same fashion. Substituting
$\hat{\psi}_i$ from (\ref{psiphi1}) into (\ref{trfa}), we find
\be\label{aphi1}
\hat{B}_i \sim e^{2A(r)} \, v_i \, \bar{\epsilon} (\gamma^\mu \partial_\mu
{f})
\, \tilde{\epsilon} \,,
\ee
and SUSY guarantees that this is a solution of (\ref{tta}). Since Killing
and anti-Killing spinors have opposite chirality this simplifies to the
form
\be\label{aphi2}
\hat{B}_i = e^{2A(r)} \, v_i \, \partial_r {f}(x^i,r) \,.
\ee
It is easy to verify that, if ${f}(x^i,r)$ satisfies (\ref{phieq}), then
$\hat{B}_i$ satisfies the massive vector equation of motion (\ref{tta})
which can be written in explicit form as
\be\label{aeq}
-(\partial_r+2A'(r))\partial_r \hat{B}_i +e^{-2A(r)}\square \hat{B}_i 
-2A''(r)\hat{B}_i=0 \,.
\ee
This completes our presentation of the TT modes of the gravity multiplet
in terms of the auxiliary scalar ${f}$.

The equations (\ref{ttgrav}), (\ref{ttpsi}), (\ref{tta}) have been
extracted from those of the full gauged ${\cal N}=8$ theory linearized
around the GPPZ flow, as can be seen explicitly from (\ref{vectact}) and
(\ref{feractke}).  Further, equation (\ref{ttgrav}) is known to be
universal for any RG flow geometry.  The gravitino equation
(\ref{ttpsi}) obtains for any reduction of the $\cN =8$ theory, since
the only other term in the gravitino field equation is proportional to
$\gamma^\mu$ and cannot contribute \cite{GRW}; it is quite
generic, since it consists merely of the canonical Rarita-Schwinger
kinetic term and a mass term.

Equation (\ref{tta}) is more particular; we will find a different
equation for the Coulomb branch graviphoton in section
\ref{CoulombSec}.  One may show, however, that given the assumption of
a graviphoton with canonical kinetic terms and some mass $m_B^2$, the
transformation (\ref{trfa}) and consistency with the gravitino
equation (\ref{ttpsi}) require $m_B^2 = -2A''$.  As we will discuss,
flows with active scalars in hypermultiplets must have canonical
vector kinetic terms, and are associated with backgrounds with broken
$U(1)_R$.  Thus it seems likely that the form (\ref{tta}) is generic
for flows where $R$-symmetry is broken, and the gauge field acquires a mass.
One should note that $m_B^2=-2A''(r)$ is non-negative, as a
consequence of the holographic $c$-theorem \cite{GPPZ1,FGPW1}.

In the Coulomb case, $R$-symmetry is preserved and the graviphoton
remains massless.  However the active scalar, which sits in a vector
multiplet, produces a non-canonical kinetic term for $B_\mu$. We will
show in section~\ref{CoulombSec}, however, that the graviphoton
equation of motion can be transformed into (\ref{aeq}), including the
same mass.  For all these reasons the vector mass $m_B^2 = - 2 A''$
appears to be generic in RG flows.

\section{The anomaly multiplet}
\label{AnomalySec}

As we have seen in the previous section, the transverse and traceless
modes of the gravity multiplet do not mix with other fields, and can
be collectively described in terms of an auxiliary free massless
scalar $f$.  The remaining modes in the gravity multiplet vanish or can be
gauged away in an anti-de Sitter background.  In an RG flow
background, however, the story is not so simple, as the profile of the
active scalar $\phi$ couples them to other fields.

The mixing of the graviton trace and the fluctuations of the active
scalar(s) was first discussed in \cite{DF}.  The equations of motion
governing the coupled system are quite general, regardless of the
character of the background flow.  The system was examined in an axial
gauge, and for the case of a single active scalar, was reduced to an
uncoupled third-order equation. This was solved for the GPPZ and
Coulomb branch flows.  Translating the solutions into sensible
correlation functions, however, proved difficult.

The problem was taken up by Arutyunov, Frolov and Theisen (AFT)
\cite{AFT}, who employed a different gauge and a different
prescription for correlation functions.  They obtained a solution for
the GPPZ flow, which we will show shortly to be gauge equivalent to
the solution of \cite{DF}.  AFT did not solve the corresponding
equation for the Coulomb branch flow, but in fact the gauge transform
of the solution from \cite{DF} satisfies their equation, as we shall
describe.  Thus the solutions are equivalent.  

In the next several sections, we shall illustrate how the coupling of
the gravity trace to the active scalar in the GPPZ flow generalizes to
a coupling between the ``trace'' of the gravity multiplet and the
active hypermultiplet.  The trace of the gravity multiplet contains
the graviton trace, gravitino $\gamma$-trace and the longitudinal
graviphoton, which are dual in the field theory to the trace of the
stress tensor, the $\gamma$-trace of the supercurrent and the
divergence of the $R$-current.  These operators constitute a chiral
multiplet called the anomaly multiplet, which vanishes when conformal
invariance is unbroken.  In the GPPZ flow background, $\gamma^\mu
\psi_\mu$ couples to the spin-1/2 fields in the active hypermultiplet,
while the phase associated to the active scalar Higgses the
graviphoton, corresponding to the breaking of the $R$-symmetry.  Thus
the coupling we uncover between the traces of the gravity multiplet
and the active hypermultiplet agrees perfectly with field theory
expectations: we can identify the multiplet of the active scalar as
the ``anomaly hypermultiplet''.

We shall examine, in turn, the graviton trace/active scalar system,
the vector/scalar system, and the fermion sector of the GPPZ flow.  We
shall see that all the anomaly multiplet fields (as well as the traces
of the gravity multiplet which mix with them) have a common spectrum
of states distinct from the transverse traceless gravity fields, while
the uncoupled Lagrangian multiplet is characterized by a third spectrum.

Interestingly, the situation is not quite the same in the Coulomb
branch flow. We will examine this case in section \ref{CoulombSec}.

\subsection{Axial and {AFT} gauges for graviton 
trace/active scalar sector}

The coupled $h_\mu^\mu / \phit$ system was considered in two
different gauges in \cite{DF} and \cite{AFT}.  Here we will
demonstrate the gauge equivalence of the solutions for the GPPZ flow,
and show that the gauge transform of the axial gauge Coulomb
branch solution solves the AFT fluctuation equation.  Finally
we will demonstrate that the effective scalar $s$ defined by AFT can be
interpreted as the active scalar $\phit$ itself in a third gauge.
We calculate the 2-point functions for the active scalar in both flows;
the result for the GPPZ flow agrees with \cite{AFT} up to the lack of
a massless pole,
while the result for the Coulomb-branch flow
displays the usual mass gap and continuum.

The most general form of the metric and scalar we use will be
\begin{eqnarray}
ds^2 &=& e^{2A(r)} \left( \eta_{ij} + h_{ij}(r,x) \right) dx^i dx^j + (-1 +
h_{55}(r,x)) dr^2 \,, \\
\phi_{tot}
 &=& \phi(r) + \phit(r,x) \,,
\end{eqnarray}
where we have already gauged away possible $h_{i5}$ components.  Since
the five TT modes of $h_{ij}$ have been discussed in
section~\ref{TTSec}, and three additional longitudinal modes can be
gauged away using (\ref{4ddiffeo}) below (see section 2.2 of
\cite{DF}), it is sufficient to restrict to the trace components of
$h_{ij}$,
\begin{eqnarray}
h_{ij}(r,x) = e^{ipx} \left( \frac{1}{4}\, h(r,p) \, \eta_{ij} + p_i
\, p_j \, H(r,p) \right) \,.
\end{eqnarray}
The holographic $\beta$-function \cite{Behr,dBVV,AGPZ} of the operator
${\cal O}_\phi$
dual to $\phi$ is defined using $a \equiv e^A$ as the scale in the
4D theory:
\begin{eqnarray}
\label{beta}
\beta_{\phi} = a \frac{d}{da} \phi = 
\frac{\phi'(r)}{A'(r)} = - \frac{3}{2W} 
{\partial W \over \partial \phi}
\,.
\end{eqnarray}
This differs by an overall sign from the $\beta$-function used by AFT.

The fields $h$, $H$, $h_{55}$ and $\phit$ all mix.  However,
residual gauge freedom can be used to eliminate one of them. In \cite{DF}
an axial gauge choice,
\begin{eqnarray}
h^{axial}_{i5} = h^{axial}_{55} = 0 \,,
\end{eqnarray}
was made, while {AFT} employed the gauge
\begin{eqnarray}
h^{AFT}_{i5} = \phit^{AFT} = 0 \,.
\end{eqnarray}
Let us consider residual diffeomorphisms defined by a vector field
$v_{\mu}(r,x)$.  In both gauges, we must require
\begin{eqnarray}
\label{ri5constraint}
\delta h_{i5} = 0 = \partial_i v_5 + \partial_5 v_i - 2 A' v_i = 0 \,,
\end{eqnarray}
which determines $v_i$ in terms of $v_5$,
\begin{eqnarray}
\label{epsiloniepsilon5}
v_i(r,x) = - e^{2A(r)} \int^r dr' e^{-2A(r')} \partial_i v_5(r',x) \,.
\end{eqnarray}
In axial gauge, we must also enforce
\begin{eqnarray}
\label{55constraint}
\delta h^{axial}_{55} = 0 = \partial_r v_5(r,x) \,.
\end{eqnarray}
While in AFT gauge, instead we have
\begin{eqnarray}
\label{phitconstraint}
\delta \phit_{AFT} = 0 = v^{\mu} \partial_{\mu} \phi(r) = -
v_5 \phi'(r) \,,
\end{eqnarray}
where $\phi(r)$ is the background scalar profile.
In both gauges, there are residual transformations of the same form as
4D diffeomorphisms:
\begin{eqnarray}
\label{4ddiffeo}
v_{i} &=& e^{2A(r)} \, w_i(x), \quad \quad v_5 = 0
\,, \\ \delta h_{ij} &=&  \partial_i \, w_j(x) +
\partial_j \, w_i(x) \,. \nonumber
\end{eqnarray}
These certainly satisfy (\ref{ri5constraint}) and
(\ref{55constraint}) or (\ref{phitconstraint}), and we may think of
them as coming from the lower limit of integration in
(\ref{epsiloniepsilon5}).

In axial gauge there are also less trivial residual gauge
transformations which are the linearization of the subgroup of bulk
diffeomorphisms which induce Weyl transformations of the boundary
metric. They have been studied in the context\footnote{DZF thanks
Kostas Skenderis for useful discussions of this point.}  of the
$AdS$/CFT correspondence in \cite{ISTY}, although their identification
is
much older \cite{PR,BH}. They are generated by
arbitrary $v_5=v_5(x)$, independent of $r$, with $v_i$ determined by
(\ref{epsiloniepsilon5}). These transformations give a pure
gauge solution \cite{DF} of the fluctuation equations in axial gauge,
namely, if $v_5(x)= e^{ipx}E(p)$, 
\begin{eqnarray}
\label{Universal}
\delta h = - 8 A'(r) \, E(p) \,, \quad \delta H' = 2 \, e^{-2A(r)} \,
E(p)\,, \quad \delta \phit = - \phi'(r) \, E(p)\,.
\end{eqnarray}
However, because $v_5$ appears without derivative in
(\ref{phitconstraint}), there is no corresponding residual
transformation in AFT gauge, which is thus a more complete gauge fix.

The fluctuation equations in axial gauge have the spurious solution
(\ref{Universal}) in addition to physical solutions, and this appears
to be the reason why these equations are more complicated and can at
best be reduced to an uncoupled third order equation for $\phit$.
In $AFT$ gauge there is no obstruction to a second order equation.

The solutions found by \cite{DF} and \cite{AFT} are nonetheless
equivalent. Consider a solution $\{ h, H, \phit \}$ in axial gauge.
Let us transform to AFT gauge by means of a vector field
$v_{\mu}(r,x)$.  We must require
\begin{eqnarray}
- \phit = \delta \phit =  - v_5 \phi'
\quad \rightarrow \quad v_5 = \frac{\phit}{\phi'} \,.
\end{eqnarray}
This transformation will introduce a nonzero $h_{55}$ and modify
$h$ and $H$.  We find
\begin{eqnarray}
\delta h = - 8 A'(r) v_5 = -8 \frac{A'(r)}{\phi'(r)} \phit =
\frac{16}{3} \frac{W \phit}{\partial W /\partial\phi} = 
- \frac{8}{\beta_{\phi}} \, \phit \,.
\end{eqnarray}
For the GPPZ flow the axial-gauge solution was \cite{DF}
\begin{eqnarray}
\label{symh}
\mt_{axial} &=& 
\frac{\sqrt{1-u}}{u} \, \left[ 4 F_1 - f_0 + p^2 L^2 u F_2 \right]
\,, \\ \nonumber
h_{axial} &=& \frac{1}{3 \sqrt{3} u} \left( 24f_0 - 96 F_1 - 24 L^2 p^2 u
F_2 +24 L^2 p^2 u (1-u) F_3 \,  \right. \\ &+& \left.  L^2 p^2 u^2 (1-u)
(8 - L^2 p^2) F_4 \right) \,, \\ \nonumber
&=& \frac{1}{3 \sqrt{3} u} \left( 24f_0 - 96 F_1 \right) \,,
\end{eqnarray}
where we have not reproduced the solution for $H$, and $f_0$ is an
integration constant associated to the pure gauge solution
(\ref{Universal}).  In the last line we used a hypergeometric
identity \cite{Bate}.  
The hypergeometric functions $F_n \, (u;p)$ are defined by
\begin{eqnarray}
F_n \, (u;p) \equiv F \left( n - \frac{3}{2} + \frac{1}{2}\, q,
n - \frac{3}{2} - \frac{1}{2}\, q; n;u \right) \,, \quad \quad
q = \sqrt{1 + p^2 L^2}\,.
\end{eqnarray}
Passing to AFT gauge, we obtain
\begin{eqnarray}
\label{deltah}
\delta h = \frac{8}{\sqrt{3}} \frac{1}{\sqrt{1-u}} \, \mt  =
\frac{8}{\sqrt{3}u} \left( -f_0 + 4 F_1 + p^2 L^2 u F_2 \right) \,.
\end{eqnarray}
giving
\begin{eqnarray}
h_{AFT} &=& \frac{8}{\sqrt{3}} p^2 L^2 F_2 \,, 
\end{eqnarray}
Up to irrelevant overall factors, this is precisely the solution
for $h$ obtained by AFT.

For the Coulomb branch flow, the solution in axial gauge was
\cite{DF}\footnote{An extraneous factor of $1/(a-1/3)$ multiplying $h$
that appeared in the first version of \cite{DF} has been corrected.}
\begin{eqnarray}
\varphit_{axial} &=& 
v^a (1-v)\ _3F_2 
\left( 1+a, 2+a, \fracs13 +a ; 2+2a, \fracs43 + a;v \right) \,, \\ 
\label{coulombh}
h_{axial} &=& \frac{4 \sqrt{2} v^a\ell^2}{3 \sqrt{3} L^4 p^2} \left[ 4
(1+3a) (2-4v-v^2 + a(2-v-v^2)) F(1+a,2+a;2+2a;v) \right. 
\nonumber\\ 
&+& 2 v(1-v)(2+v) (2+a)(1+3a) F(2+a,3+a;3+2a;v)
\\   
&+&3  \left. (2+v) {L^4 p^2\over \ell^2} \: _3F_2 \left( 1+a, 2+a,
\frac{1}{3} +a; 2+2a, \frac{4}{3} +a; v \right) \right]  \,, \nonumber 
\end{eqnarray}
where $_mF_n(b_1,...,b_m; c_1,...,c_n; z)$ denotes a 
generalized hypergeometric function \cite{Bate} and
$a~\equiv~(-1 + \sqrt{1 - p^2 L^4 / \ell^2})/2$.  Transforming
into {AFT} gauge gives
\begin{eqnarray}
\delta h &=& - \frac{4 \sqrt{6}}{3} \frac{v+2}{1-v} \, \phit = - \frac{4
\sqrt{6}}{3} v^a (v+2) \, _3F_2 \left( 1+a,
2+a, \frac{1}{3} +a; 2+2a, \frac{4}{3} +a; v \right)  \\
\label{coulhAFT}
h_{AFT} &=& - \frac{4 \sqrt{2} \ell^2 v^a}{3 \sqrt{3} L^4 p^2} (1+3a) 
\left\{ 4 [-2+4v+v^2 + a(-2+v+v^2)] \; \times \right. \\ && \left.
F(1+a,2+a;2+2a;v) - 2  v(1-v)(2+v) (2+a) F(2+a,3+a;3+2a;v)\right\}
\nonumber \,,
\end{eqnarray}
where the unpleasant $_3F_2$ hypergeometric function exactly cancels.
The solution (\ref{coulhAFT}) satisfies the AFT equation of motion,
\cite{AFT} equation (2.32), where $s$ is related to $h_{AFT}$ by
(\ref{hands}).

\subsection{Dynamical scalar gauge}
\label{DynamicalSec}
 
Finally, we wish to discuss a third gauge, where the active scalar is
kept as the dynamical degree of freedom.  AFT claim that by
extracting an $r$-dependent factor from $h$, one is left with a field
$s$ which is the scalar coupling to ${\cal O}_{\phi}$:
\begin{eqnarray}
\label{hands}
h_{AFT} = - \frac{8}{\beta_{\phi}} \, s \,,
\end{eqnarray}
where we added the minus sign to accord with our definition of the
$\beta$-function (\ref{beta}).  One can show explicitly that this
identification is correct by transforming from AFT gauge, where
$\phit=0$, to a gauge where $h= 0$.  We must require
\begin{eqnarray}
- h = \delta h = - 8 A'(r) v_5 \quad \quad \rightarrow \quad \quad 
v_5 = \frac{h}{8 A'(r)} \,.
\end{eqnarray}
Then
\begin{eqnarray}
\phit_{DS} = - v_5 \phi'(r) = - \frac{\phi'(r)}{8 A'(r)} 
\, h_{AFT} = \frac{3}{16W} \frac{\partial W}{\partial\phi} h_{AFT}  = -
\frac{\beta_{\phi}}{8} h_{AFT} \,.
\end{eqnarray}
and by virtue of (\ref{hands}),
\begin{eqnarray}
\phit_{DS} = s \,.
\end{eqnarray}
This third gauge, in which the dynamics of the system is carried by
the active scalar, is perhaps the most intuitive; we will use
something similar in the fermion sector.  It also seems likely that
this gauge will generalize most easily to the case of several active
scalars, since it involves a condition on $h$ instead of on the set of
scalars.

Using the well-established procedure for calculating 2-point functions
of operators dual to scalar fields \cite{holo,FMMR}, we find for the
GPPZ case
\begin{eqnarray}
\langle{\cal O}_{\mt}(p) {\cal O}_{\mt}(-p) \rangle = \frac{p^2}{2} \left[
\psi \left( \frac{3}{2} + \frac{1}{2} \sqrt{1 + p^2 L^2} \right) +
\psi \left( \frac{3}{2} - \frac{1}{2} \sqrt{1 + p^2 L^2} \right) \right] \,,
\end{eqnarray}
which agrees with \cite{AFT}, eqn.~(2.26) up to normalization.  The
apparent difference in the arguments of the $\psi$ functions only
corresponds to the addition of a contact term.  The spectrum is $p^2 =
4(n+1)(n+2)/L^2$, which will prove to be common to all members of the
anomaly multiplet, as in equation (\ref{Aspec}).  Note that there is
no massless pole.

For the Coulomb branch case, the active scalar corresponds to a
$\Delta =2$ operator, and so we must use the modified prescription for
the 2-point function, see \cite{MR}.  We obtain
\begin{eqnarray}
\langle{\cal O}_{\varphit}(p) \, {\cal O}_{\varphit} (-p) \rangle = -
\lim_{\epsilon \rightarrow 0} \, \left( \epsilon^4 \log^2 \epsilon
\right) \left[ \frac{1}{z^3} \frac{d}{dz} \ln(\varphit(z,p))
\right]_{z= \epsilon} = \psi \left( \frac{1}{2} + \frac{1}{2} \sqrt{1
- \frac{p^2 L^4}{\ell^2}} \right) \,,
\end{eqnarray}
with the $z$-variable defined in \cite{DF}.  This functional form is
ubiquitous for Coulomb branch 2-point functions, and indicates the
mass gap at $m^2_{gap} = \ell^2/L^4$ and overlying continuum.

Returning to general considerations, one may notice that although the
boundary scaling of $\phit$ is the same as that of any scalar field
acting as a source, $\phit \sim e^{(\Delta-4)r}$, the behavior of
the field $h$ in AFT gauge is somewhat unusual.  If the background
corresponds to an operator deformation of ${\cal N}=4$ SYM, then the
$\beta$-function scales as $\beta \sim e^{(\Delta-4)r}$, and we have
\begin{eqnarray}
h_{AFT} \sim const \,,
\end{eqnarray}
near the boundary.  On the other hand, if the background corresponds
to a different vacuum, $\beta \sim e^{-\Delta r}$ and we find for
$\Delta > 2$,
\begin{eqnarray}
\label{couldiv}
h_{AFT} \sim e^{(2 \Delta - 4)r} \,,
\end{eqnarray}
which diverges on the boundary.\footnote{For the special case of
$\Delta = 2$, we have $h \sim r$ for a Coulomb background, which is
also divergent.}  Thus fluctuations of $h$ are constant in an operator
background, but diverge for the vev background.

In the Coulomb branch, the stress-tensor trace $T^\mu_\mu =0$
vanishes, rather than being determined by the operator ${\cal
O_\phi}$ as it is on an operator flow. 
The relation between this vanishing and the divergent behavior
of (\ref{couldiv}) is not clear to us.

\section{The vector/scalar sector}
\label{ScalarSec}

This section is devoted to studying the spectrum of $SO(3)$-invariant
scalar fluctuations around the GPPZ flow \cite{GPPZ} with $\sigma = 0$
in the background.  As discussed in section~\ref{InvariantSec}, the
eight real scalars, denoted $m$, $\sigma$ and $t_{(ij)}$ in the
Iwasawa/solvable parameterization, describe the quaternionic coset
$G_{2(2)}/SO(4)$.

The active scalar $m$ and its ``imaginary partner'' $t_{(10)}$ sit in
the active hypermultiplet.  They are dual to the $F$-component of the
anomaly multiplet ${\cal A} = {\Tr}(\Phi^i)^2$ in the ${\cal N}=1$
superfield description of the mass-deformed field theory, namely the
operator ${\Tr}({1\over 2}\psi_i \psi_i + z_i \, F_i)$ 
of dimension $\Delta=3$.  The
bulk fields $m$ and $t_{10}$ mix with the (trace of the) graviton and
the (longitudinal component of the) graviphoton,
respectively.\footnote{As we will see momentarily, it would be more
appropriate to say that the ``modulus" $\chi$ of the complex active
scalar mixes with the graviton and its ``phase" $\beta$ mixes with the
graviphoton.}  The scalar $\sigma$ and its imaginary partner
$t_{(32)}$ reside in the dilaton hypermultiplet, and are dual to the
$\Delta = 3$ gaugino bilinear operator in the lowest component of the
Lagrangian multiplet ${\cal S} = {\Tr} (W_\alpha W^\alpha)+ \ldots$

The remaining four scalar singlets mix in pairs, $t_{(11)}/
t_{(31)}$ and $t_{(21)}/t_{(01)}$, but do not mix with other
fields.  We will exploit SUSY to diagonalize these fields, and we find
that each pair contributes a real scalar dual to a $\Delta = 2$
operator in the active hypermultiplet, and a real scalar dual to a $\Delta =
4$
operator in the dilaton multiplet.  The two $\Delta = 4$ fields are the 5D
axion/dilaton $\tau$, whose kinetic term is non-canonical and whose
$AdS$ mass is only asymptotically vanishing near the boundary.  Note
that there is no physical ``massless" scalar field whose spectrum of
fluctuations would coincide with that of the transverse components of
the traceless graviton and graviphoton.

We will first consider the scalars that do not mix with the graviton
and the graviphoton. For our purposes we need the quadratic scalar
Lagrangian and the linearized SUSY transformation rules. Two of the
scalars, $\sigma$ and $t_{(32)}$, have diagonal, canonical kinetic
terms; the equation for $\sigma$ has already been solved in \cite{DF}
and $t_{(32)}$ is handled here analogously.  In order to diagonalize
the equations governing the fluctuations of $t_{(11)}/t_{(31)}$
and $t_{(21)}/t_{(01)}$, we rely on an {\it ansatz} suggested
by the preserved supersymmetry.  We then resolve the
vector/pseudoscalar mixing by performing a field redefinition that
brings the relevant Lagrangian into St\"uckelberg form.  The
graviton/active scalar mixing has been discussed at length in section
5.

\subsection{Free scalar equations of motion}

The quadratic Lagrangian for the scalar fields mixing only among 
themselves can be compactly written in the form
\be
{1\over \sqrt{g}}{\cal L} = {1\over 2} G_{IJ}(m)  \partial \phit^{I} 
\partial 
\phit^{J} -  {1\over 2} M^{2}_{IJ}(m) \phit^{I} \phit^{J} \quad ,
\ee
where as always $\phit^{I}$ denote the fluctuations around the classical
solution.
The symmetric tensors $G_{IJ}$ and $M^{2}_{IJ}$ are functions of $r$ though
their dependence on the active scalar $m$.
Looking for solutions of the form,
\be
\phit^{I} (r, x) = \phit^{I} (r, p) e^{-i p\cdot x} \,,
\ee
we obtain in the RG flow background the fluctuation equations
\be
\partial_{r}^{2} \phit^{I} + 4 A' \partial_{r} \phit^{I} +
T^{I}_{rJ}\partial_{r} \phit^{J} + e^{-2A} p^{2}
\phit^{I} - Z^{I}{}_{J} \phit^{J} = 0 \,,
\ee
where $T^{I}_{rJ} \equiv G^{IK}\partial_{r}G_{JK}$ and
$Z^{I}{}_{J} \equiv G^{IK}M^{2}_{JK}$.

As is the case with other fluctuation equations in this background, it
is convenient to switch to the radial variable $u$ (\ref{udef}). One finds
\be
u(1-u) \partial_{u}^{2} \phit^{K}  +
\left[(2-u) \delta^{K}{}_{J} + {u\over 1-u} T^{K}_{uJ}
\right]
\partial_{u} \phit^{J} 
+ \left[ {p^2L^2 \over 4} \delta^{K}{}_{J} 
- {L^2 \over 4}{u\over 1-u} Z^{K}_{J}
\right]
\phit^{J} = 0 \quad .
\label{uscalareom}
\ee
We now apply equation (\ref{uscalareom}) to each of the scalar sectors.

The fields $\sigma$ and $t_{(32)}$ appear diagonally in the quadratic 
Lagrangian.\footnote{For GPPZ flows with $\sigma\neq 0$ the situation is 
slightly more involved. $\sigma$ has to be treated at the full 
non-linear level since it mixes with the graviton. $t_{(32)}$ or better 
the phase of the {\it a priori} complex $\sigma$ mixes with the 
graviphoton.}
They have canonical metric $G_{\sigma\sigma} = G_{(32)(32)} = 1$ 
and the same mass term, 
\be
M^{2}_{\sigma\sigma} = M^{2}_{(32)(32)} = {3 \over L^{2}} 
\left[1-2 \cosh \left(\frac{2 m}{\sqrt{3}} \right) \right] =
{3 \over L^{2}} \left( {3 u - 4 \over u}\right ) \,,
\ee
and thus have an identical fluctuation equation,
\be
u(1-u) \sigma'' + (2-u) \sigma' + \left({p^2L^2 \over 4} -  {3 (3u- 4) \over
4 (1-u)}  \right) \sigma = 0 \,.
\ee
which was solved in \cite{DF}, giving
\begin{eqnarray}
\{\sigma, t_{(32)} \} = (1-u)^{3/2} \, F \left( \frac{3}{2} +
\frac{1}{2} \sqrt{9 + p^2L^2} , \frac{3}{2} - 
\frac{1}{2} \sqrt{9 + p^2L^2}; 2; u \right) \,.
\end{eqnarray}
This function has poles where ${3}\pm \sqrt{9 + p^2L^2} = -2n$, $n$
integer.  
The
spectrum of poles is $p^2L^2 = 4n(n+3)$, as in equation (\ref{Sspec}),
including a massless state.

The scalars $t_{(11)}$, $t_{(31)}$, $t_{(21)}$ and $t_{(01)}$ form two
$2\times 2$ independent sectors with diagonal non-canonical metric and
non-diagonal mass matrices, as given in equations (\ref{scalarkin}),
(\ref{quadpot}).  For compactness of notation we often put $\mu = 2 m
/ \sqrt{3}$ in the following.

In the $t_{(11)}/t_{(31)}$ sector, it is convenient to put $ \phit^{1}_+
\equiv t_{(11)} $ and $ \phit^{2}_+ \equiv 
t_{(31)} $; similarly, in the  $t_{(21)}/t_{(01)}$ sector, we define $
\phit^{1}_- \equiv t_{(21)} $ and $ \phit^{2}_- \equiv t_{(01)} $.
One then finds the functions
\ba
&G_{11} =  e^{ \pm \mu} \qquad
&M^{2}_{11} = {1\over 4} (-10 + 11 e^{ \pm \mu} - 6 e^{ \pm 2\mu} + e^{ \pm
3\mu}) 
\nonumber \\
&G_{22} =  e^{ \mp 3\mu} \qquad
&M^{2}_{22} = {1\over 4} (-3 e^{\mp \mu} - 12 e^{\mp 2\mu} + 3 e^{\mp 3\mu})
\nonumber \\
&G_{12} =  0 \qquad
&M^{2}_{12} = {\sqrt{3}\over 4} (1 - 6 e^{\mp \mu} + e^{\mp 2\mu}) \quad .
\label{mixedGM}
\ea
It is remarkable that these two
sectors are simply related by a change of sign of $\mu$.
Thus, performing the analysis in one of the two sectors, say  
$t_{(11)}/t_{(31)}$,
suffices for both.
The equations of motion are then of the form
\be
\label{reducedeqn}
\phit^{I}{}'' + (4 A' + a_{I}) \phit^{I}{}'
+ e^{-2A} p^{2}\phit^{I} - b_{I} \phit^{I} - c_{I}
\phit^{\bar{I}} = 0 
\ee
where
$a_{I} \equiv d \log(G_{II}) /dr $, $b_{I} \equiv (G_{II})^{-1} M_{II}$ and
$c_{I} \equiv (G_{II})^{-1} M_{I\bar{I}}$ with $\bar{I} = I+1 ({\rm mod}\ 2)$.
One can easily get an uncoupled fourth-order equation for one of the scalars
by eliminating its partner.  However, 
exploiting the linearized SUSY preserved by the kink solution, we
can deduce an ansatz for the scalar fluctuations that turns out to reduce
the problem to standard second-order ordinary differential 
equation.

Consider the SUSY transformations involving $t_{(11)}$ and $t_{(31)}$
in (\ref{susyone}), (\ref{susytwo}). Their form strongly suggests that
it is the separate linear combinations, whose transforms involve
$\xi^{1,2}$ and $\xi^{3,4}$ respectively, which satisfy uncoupled
scalar equations. A similar observation can be made for $t_{(21)}$ and
$t_{(01)}$.  This leads us to define 
\ba
\phit^{1}_\pm &\equiv&  e^{\mp {\mu/2}} ( \rho_\pm - \sqrt{3} \tau_\pm ) \,,
\nonumber \\
\phit^{2}_\pm &\equiv& e^{\pm {3\mu/2}} ( \sqrt{3}  \rho_\pm + \tau_\pm )
\,.
\label{susyansatze}
\ea
Indeed we will find that $\rho$ and $\tau$ decouple and satisfy
second order equations whose mass terms $(mL)^2$ approach $-4$ and 0,
respectively, as $r \rightarrow \infty$. Thus $\rho,\tau$ are dual
to operators of dimension $\Delta = 2,4$ respectively.

To verify decoupling we substitute the combinations (\ref{susyansatze}) into
the
fluctuation equations (\ref{reducedeqn}) and {\it assume} that the resulting
equations are satisfied independently by $\rho$ and $\tau$. One
then finds two second order equations for each of the four
modes $\rho_\pm,\tau_\pm$ with complicated and apparently
different effective mass terms. Miraculously, when the background
flow equations (\ref{gravkilling}), (\ref{chikilling2}) and the GPPZ
superpotential (\ref{gppzW}) are used,
the two equations are seen to be equivalent, which {\it proves} the
consistency of the assumed decoupling. A further simplification is
that the equations for the $\pm$ modes are even in $\mu$ and thus identical!
For $\rho=\rho_+=\rho_-$ the final fluctuation equation takes the
relatively simple form,
\be
\rho'' + 4 A' \rho' +
\left( e^{-2A} p^{2} + {1 \over 4L^{2}} \left[9 + 10 \cosh(\mu) - 3
\cosh(\mu)^{2} \right] \right)  \rho = 0 \,.
\ee
In the variable $u$ this becomes
\be
u(1-u) \rho'' + (2-u) \rho' + \left({p^2L^2 \over 4}
 -  {u^{2} - 8 u +3 \over 4u(1-u)}
\right) \rho = 0 \,,
\ee
which has the solution
\begin{eqnarray}
\rho = u^{1/2} (1-u) \, F \left( \frac{3}{2} + \frac{1}{2} \sqrt{1 +
p^2L^2}, \frac{3}{2} - \frac{1}{2} \sqrt{1 + p^2L^2};3;u \right) \,,
\end{eqnarray}
The spectrum of poles is at ${3}\pm \sqrt{1 + p^2L^2} = 
-2n$, giving $(pL)^{2} =
4(n+1)(n+2)$, as appropriate for the multiplet ${\cal A}$,
see equation (\ref{Aspec}).  

The active scalar $\mt$, reviewed in section~\ref{AnomalySec}, and its
partner $t_{(10)}$, which we examine in the next subsection, share
this spectrum of poles with $\rho_+$ and $\rho_-$, leading to a
fourfold degeneracy.  This is exactly what one expects on SUSY
grounds, as $m$ and $t_{(10)}$ are dual to the top component $F_{\cal
A}$ of the anomaly multiplet ${\cal A}$, with $\Delta = 3$, while
$\rho_\pm$ are dual to the lowest component $\phi_{\cal A}$ with
$\Delta = 2$.  The mixing of $\mt$ and $t_{(10)}$ with
$h^\mu_\mu$ and $\partial^\mu B_\mu$ is dual to the fact that $F_{\cal
A}$ contains $T^\mu_\mu$ and $\partial^\mu R_\mu$, while the fact that
$\rho_\pm$ do not mix with the gravity multiplet is appropriate since
$\phi_{\cal A}$ does not correspond to any mode of the stress tensor
or $R$-current.  The fermionic components of this multiplet, 
$\xi^1$ and $\xi^2$, will be shown to have the same spectrum
in section~\ref{FermionSec}.

For the fields $\tau = \tau_{+} = \tau_{-}$ one obtains equations equivalent
to 
\be
\tau'' + 4 A' \tau' + \left( e^{-2A} p^2 - {1 \over 4L^{2}}
\left[15 - 18 \cosh(\mu) + 3 \cosh(\mu)^2 \right] \right) \tau = 0 \,,
\ee
which, thanks to the symmetry between the two ($+$ and $-$) sectors, 
is even in $\mu$. In the $u$-variable one gets
\be
u(1-u) \tau'' + (2-u) \tau' + \left({p^2L^2 \over 4} +  { 3 (3 u - 1) \over 4u}
\right) \tau = 0 \,,
\ee
which is solved by
\begin{eqnarray}
\tau = u^{1/2} (1-u)^2 F \left(\frac{5}{2} + \frac{1}{2} \sqrt{9 + p^2L^2},
\frac{5}{2} - \frac{1}{2} \sqrt{9 + p^2L^2};3;u \right) \,.
\end{eqnarray}
The spectrum of poles is $(pL)^{2} = 4(n+1)(n+4)$, identical
to that of the scalar fields $t_{(32)}$ and $\sigma$ except for 
the absence of
the zero-mass pole. 
Indeed the four scalars in question are dual to the two complex
scalar components of the
chiral multiplet ${\cal S} = {\Tr} (W^{2}) + \ldots$. The scalar fields
$\sigma$ and $t_{(32)}$ are dual to the lowest component $\phi_{\cal
S}$ with $\Delta = 3$, while the fluctuations
$\tau_\pm$ have asymptotic mass $m^2L^2 = 0$ and  are dual to
the top component $F_{\cal S}$ with $\Delta = 4$.

The fact that the massive poles are fourfold degenerate while the massless
pole is only doubly degenerate is explained by the following 
supersymmetry argument.
The two-point function of a gauge-invariant chiral superfield in 
coordinate superspace $(x, \theta, \bar\theta)$ is fixed by ${\cal N} 
= 1$ supersymmetry to be of the form
\be
\langle {\cal S}(x_1, \theta_1, \bar\theta_1) {\cal S}^{\dagger}(x_2, 
\theta_2, \bar\theta_2) \rangle = e^{i(\theta_1 \sigma \bar\theta_1 + 
\theta_2 \sigma \bar\theta_2
- 2 \theta_1 \sigma \bar\theta_2) \cdot \partial_1} \Delta (x_{12}) \,,
\ee
where $\Delta (x_{12})$ is an {\it a priori} arbitrary scalar function of the 
relative position $x^\mu_{12} = x^\mu_1 - x^\mu_2$.
After Fourier transforming and expanding in powers of $\theta$ one 
easily obtains the ``SUSY Ward identity"
\be
\langle F_{\cal S}(p) F^{\dagger}_{\cal S}(-p) \rangle =
p^{2} \langle \phi_{\cal S}(p) \phi^{\dagger}_{\cal S}(-p) \rangle \,.
\ee
We thus see that any potential simple pole at $p^2=0$ for the lowest 
component $\phi_{\cal S}$
is cancelled by the factor of $p^2$ for the top component $F_{\cal 
S}$.\footnote{Notice that $F_{\cal S}$ is auxiliary/non-propagating, \ie 
has $\delta$-function propagator in coordinate space, only for a free 
chiral superfield.}
As we will see in
section~\ref{FermionSec}, the multiplet is completed by the addition of the
fermionic 
partners $\xi^3, \xi^4$, which decouple from the
gravitino and present the expected spectrum,
including the massless pole for one of the two chiralities.

\subsection{Resolving the vector/scalar mixing}

Let us now discuss the mixing of the Goldstone field $t_{(10)}$ with
the graviphoton field $B_\mu$. The key observation here is that the four
scalar fields  $\{m, t_{(10)}, \sigma, t_{(32)} \}$  parameterize an
$SL(2)/U(1)\times SL(2)/U(1)$  submanifold of $G_{2(+2)} /SO(4)$
\cite{FS,BC} and the $U(1)_R$ generator defined in
 (\ref{uonegen})   is a linear combination of the compact $U(1)$  isometries
of the two $SL(2)/U(1)$ factors. Since we are studying a flow with
$\sigma(r)=0$, there are no bilinear mixing terms from the second
$SL(2)/U(1)$ factor, and we can restrict attention to the first factor
only. We can derive all the information we need from a standard 
Lagrangian for the gauged $SL(2)/U(1)$ $\sigma$-model,
\be
{1\over \sqrt{g}} {\cal L}_{gauged} = -{3\over4} F^2 +
{3\over 8} \left[ \partial\chi^{2} + 
\sinh^2 (\chi) (\partial\beta - g B)^{2} \right] \,,
\label{gauged}
\ee
in which $\beta$ is the angular variable of the compact $U(1)$ isometry.

We first demonstrate the equivalence of (\ref{gauged}) to the form
given in section~\ref{InvariantSec}.
The change of coordinates  
\be
\cosh(\chi)= {X^2 +Y^2 +1\over 2Y} \,,\quad
\tan(\beta) = {{2X}\over {1-X^2-Y^2}} \,,
\ee
takes us to the Poincar\'e plane form
\be
{1\over \sqrt{g}} {\cal L}_{gauged} = -{3\over 4} F^2 +
{3\over 8 Y^2} \left[ (\partial Y - g B X Y)^{2} + 
[\partial X + {g\over 2} B(1- Y^2 - X^2)]^{2} \right] \, .
\label{gaugedpoin}
\ee The further transformation \be Y=e^{2m\over \sqrt{3}} \quad X= {2
\over \sqrt{3}} t_{(10)} \ee gives a kinetic Lagrangian which agrees
to quadratic order with (\ref{scalarkin}),
(\ref{vectact}), (\ref{vectsccop}) of section~\ref{InvariantSec} (when
$\sigma =0$), but is a fully nonlinear extension thereof.

The $U(1)$ is a subgroup of the gauged $SO(6)$ invariance of the
${\cal N}=8$ supergravity theory, so the scalar potential is $U(1)$
invariant and thus independent of the field $\beta$. 
We can therefore use the $\chi$, $\beta$ form of the Lagrangian (\ref{gauged})
for our present purpose of determining the vector/scalar
fluctuations, and we can replace $\chi$ by its background
value  $\chi = \mu \equiv  2m/\sqrt{3}$. This gives
\be
\sinh^2(\chi) = {1\over 2} \left[\cosh \left({4m(r)\over \sqrt{3}} \right)-1
\right]
         = - {8 A''(r) \over g^2} \,.
\ee
The Lagrangian (\ref{gauged}) to quadratic order in fluctuations of 
$B_\mu$ and $\beta$ then becomes
\be
{1\over \sqrt{g}}{\cal L} = -{3\over 4} F^{2}
+ {3\over 2g^2} \, m_B^2  \, (\partial \beta - g B)^{2} \,,
\ee
where  $m_B^2 = -2 A''(r)$ is the graviphoton mass obtained in
section~\ref{InvariantSec} and
we have kept the unconventional normalization factor of 3 that emerged
from the truncation of ${\cal N}=8$ supergravity in that section. We
have thus reduced the system to a standard St\"uckelberg Lagrangian,
a fact which will help immensely as we now turn to the solution of its
equations of motion,
\begin{eqnarray}
\label{Stuckveceom}
D_{\mu} F^{\mu\nu} -  m_B^2/g (\partial^{\nu}\beta - g B^{\nu}) 
&=& 0 \,, \\
{1\over \sqrt{g}} \partial_{\mu} (\sqrt{g} g^{\mu\nu}
 m_B^2 (\partial_{\nu}\beta - g B_{\nu})) &=& 0 \,.
\label{psscalareom}
\end{eqnarray}
The contracted Bianchi identity for (\ref{Stuckveceom}), 
\ie current conservation,
\be
D_{\mu}D_{\nu}F^{\mu\nu} = 0 = D_{\mu} J^{\mu} \,,
\ee
implies the scalar equation (\ref{psscalareom}).

The equations are gauge invariant and we discuss their solution in two
gauges. The fastest route to a solution employs the gauge $\beta(x,r)=0$
which is the analogue of the AFT gauge of section~\ref{AnomalySec}. This is
equivalent to the St\"uckelberg approach, where one works in terms of a gauge
invariant vector field,
\be
{\cal B}_{\mu} = B_{\mu } - {1\over g} \partial_{\mu} \beta \,.
\ee
The transverse components, which were treated in section~\ref{TTSec}, 
decouple from the remaining longitudinal
and radial components,
and the latter are related by the current conservation condition
\be
D_{\mu} ( m_B^2 {\cal B}^{\mu} ) = 0 
\ee
or, more explicitly,
\be
 \eta^{ij} \partial_{i} {\cal B}_{j} = {1\over e^{2A}  m_B^2}
\partial_{r} (e^{4A}  m_B^2 {\cal B}_{r}) \,.
\label{BtoB}
\ee
We use this to write radial component of the vector equation of motion as
\be
\square {\cal B}_r - 
\partial_r \left( {1\over e^{2A}  m_B^2} 
\partial_r ( e^{4A}  m_B^2 {\cal B}_r )
\right) +  
 e^{2A}  m_B^2 {\cal B}_r = 0 \,.
\ee
One may then show that the redefined field 
$C(u,p) \equiv e^{4A(r)} m_B^2 {\cal B}_r(r,p)$ satisfies 
\be
u(1-u) C'' + (1-2u) C' + \left({p^2L^2 \over 4} - {1\over u} \right) C = 0 \,,
\ee
with the hypergeometric solution, $C(u,p) = u F_3(u,p)$ in the notation
of (90). This has the same spectrum of poles as its partner, $h_{AFT}$ or
$s$, in the anomaly multiplet. 

We now briefly discuss the axial gauge $B_r(r,x)=0$ which has  
residual gauge transformations generated by a gauge parameter $\alpha(x)$
which is independent of $r$ but otherwise arbitrary.  
In this gauge the radial component of the vector equations of motion 
implies
\be
\partial_{r} (\partial_{i} B^{i}) = 
\frac{1}{g} \, e^{2A} \, m_B^2 \, \partial_{r} \beta \,.
\label{Btobeta}
\ee
In analogy with section~\ref{AnomalySec}, one can use this to obtain the
uncoupled third
order scalar equation
\be
\partial_{r} \left( {1\over  m_B^2 e^{2A}} \partial_{r} (e^{4A}  m_B^2
\partial_{r}
\beta ) \right) = - p^{2} \partial_{r}\beta +  m_B^2 e^{2A} \partial_{r} \beta
\,.
\ee
This can be viewed as a second order equation for $\partial_r \beta$, and
one can use (\ref{Btobeta}) and (\ref{BtoB}) 
to express its solution in terms of $C(u,p)$.

Including the second $SL(2)/U(1)$ factor is an easy task. To the
Lagrangian (\ref{gaugedpoin}) one simply has to add 
\be
{1\over \sqrt{g}} \Delta {\cal L}_{gauged} =
{1\over 8 U^2} \left[ (\partial U + 3g B U V)^{2} + 
[\partial V - {3g\over 2} B(1- U^2 - V^2)]^{2} \right] \,,
\label{gaugedinert}
\ee
where
\be
U = e^{-2 \sigma} \,, \quad V = {2} t_{(32)} \, .
\ee
As remarked in section~\ref{InvariantSec}, $m_B^2 = - 2 A''$ continues to hold
even when $\sigma\neq 0$. The different normalization of the
$U(1)_R$ charge in the $\{ \sigma, t_{(32)} \}$ sector is as required by 
the generator in (\ref{uonegen}).

\section{The fermion sector}
\label{FermionSec}

We now consider the fermion sector in the GPPZ flow.  In analogy to
the coupling of the graviton trace and longitudinal graviphoton with
the active scalar's modulus and phase, respectively, we will find that
the $\gamma$-trace of the gravitino couples to the fermi fields from
the active hypermultiplet.  The spinors from the dilaton hypermultiplet
are uncoupled and display the expected spectrum.

Setting $\sigma = 0$, the fermion Lagrangian is
\begin{eqnarray}
e^{-1} {\cal L} &=& -\frac{i}{2} \left( \bar\psi^1_\mu \gamma^{\mu \nu
\rho} D_\nu \psi^2_\rho - \bar\psi^2_\mu \gamma^{\mu \nu \rho} D_\nu
\psi^1_\rho \right) - \frac{ig}{4} W \left( \bar\psi^1_\mu \gamma^{\mu
\nu} \psi^1_\nu + \bar\psi^2_\mu \gamma^{\mu \nu} \psi^2_\nu \right)
+\nonumber \\ \label{lag}
 && -\frac{i}{2} \left( \bar\xi^1 \gamma^{\mu} D_\mu
\xi^2 - \bar\xi^2 \gamma^{\mu} D_\mu \xi^1 + \bar\xi^3
\gamma^{\mu} D_\mu \xi^4 - \bar\xi^4 \gamma^{\mu} D_\mu \xi^3
\right) +\\ &&
- \frac{i}{2} M(r) \left(\bar\xi^1 \xi^1 + \bar\xi^2 \xi^2 -
3\bar\xi^3 \xi^3 - 3\bar\xi^4 \xi^4 \right)
+ \nonumber \\
&& \frac{m'(r)}{\sqrt{2}} \left( \bar\psi^1_\mu \gamma^r
\gamma^{\mu} \xi^1 + \bar\psi^2_\mu \gamma^r \gamma^{\mu} \xi^2 \right)
- \frac{\sqrt{3} g}{4 \sqrt{2}} \sinh  \left( \frac{2m}{\sqrt{3}} \right)
\left(\bar\xi^2 \gamma^{\mu} \psi^1_\mu - \bar\xi^1 \gamma^\mu \psi^2_\mu
\right) \,, \nonumber
\end{eqnarray}
where
\begin{eqnarray}
\label{Mdef}
M(r) \equiv - \frac{3}{2} A' - \frac{m''}{m'} =\frac{g}{8} \left[
\cosh \left( \frac{2m}{\sqrt{3}} \right) - 3 \right]\,.
\end{eqnarray}
The $\delta \chi^{abc} = 0$ condition for unbroken supersymmetry also
requires
\begin{eqnarray}
\label{msinhid}
{m'(r)} = - \frac{\sqrt{3} g}{4} \sinh  \left( \frac{2m}{\sqrt{3}} \right) \,,
\end{eqnarray}
which leads to the appearance of chirality projectors in the equations
of motion.  Defining the complex spinors
\begin{eqnarray}
\label{dirac}
\psi_\mu \equiv \psi_\mu^1 + i \psi_\mu^2 \,, \quad \quad
\xi \equiv \xi^1 + i \xi^2 \,,  \quad \quad
\eta \equiv \xi^3 + i \xi^4 \,,
\end{eqnarray}
one obtains the field equations
\begin{eqnarray}
\label{psieqn}
i \gamma^{\mu \nu \rho} D_\nu \psi_\rho &=& \frac{g}{2} W \gamma^{\mu \nu}
\psi_\nu + \frac{m'}{\sqrt{2}} \left( 1 + i \gamma^r \right) \gamma^\mu \xi
\,, \\
\label{xieqn}
i \gamma^\mu D_\mu \xi &=&  M(r) \xi - \frac{m'}{\sqrt{2}}
\gamma^\mu \left( 1 - i \gamma^r \right) \psi_\mu \,, \\
\label{etaeqn}
i \gamma^\mu D_\mu \eta &=& - 3 M(r) \eta \, ,
\end{eqnarray}
where here $D_i = \partial_i - {1 \over 2} A' \gamma_i \gamma_r$ and
$D_5 = \partial_5$; Christoffel connections cancel in (\ref{psieqn})
due to antisymmetry.

\subsection{The uncoupled spinor}

We begin by considering equation (\ref{etaeqn}) for the uncoupled
spinor field $\eta$.  These techniques will generalize to the coupled
$\psi_\mu / \xi$ system.  The $\eta$ field sits in the inert
hypermultiplet with the scalars $\sigma$ and the ``dilaton'' $\tau$,
and so should have the same spectrum.

Define the ``chirality'' projectors
\begin{eqnarray}
P_\pm \equiv \frac{1}{2} \left( 1 \pm i \gamma^r \right) \,, 
\end{eqnarray}
which obey
\begin{eqnarray}
P_\pm \gamma_r = \gamma_r P_\pm \,, \quad \quad
P_\pm \gamma_i = \gamma_i P_\mp \,, \quad \quad
P_\pm D_\mu = D_\mu P_\pm \,. 
\end{eqnarray}
The chiral projections of equation (\ref{etaeqn}) are then
\begin{eqnarray}
\label{proj1}
i \dslash \eta_+ -  2 A' \eta_-
+ i \gamma^r \partial_r \eta_- + 3 M(r) \eta_- = 0 \,, \\
\label{proj2}
i \dslash \eta_- +  2  A' \eta_+
+ i \gamma^r \partial_r \eta_+ + 3 M(r) \eta_+ = 0 \,, 
\end{eqnarray}
where $\dslash \equiv \gamma^i \partial_i$.  Our strategy is to eliminate one
of the projections of $\eta$ and solve for the other.  
Writing $\eta(x,r) = e^{ipx} \eta(p,r)$, we can solve for $\eta_-$ using
(\ref{proj2}),
\begin{eqnarray}
\label{solvetaminus}
\eta_- = \frac{1}{\pslash} \left( \partial_r + 2 A' + 3M \right) \eta_+ \,. 
\end{eqnarray} 
We then substitute (\ref{solvetaminus}) into (\ref{proj1}). The
identity $\partial_r (1/\!\!\!\!\pslash) = A'/ \!\!\pslash$ is needed.
Finally we multiply by an overall $\pslash$ and use $\pslash \!
\pslash = g^{ij} p_i p_j$ to obtain
\begin{eqnarray}
\label{etapluseqn}
\left(- e^{-2A} p^2 - \partial_r^2 - 5 A' \partial_r - 6 A'^2 - 2 A'' + 9 M^2
- 3 M A' - 3 M' \right) \eta_+ = 0 \,,
\end{eqnarray}
where as usual $p^2 = \eta^{ij} p_i p_j$.  Thus we find the same
ordinary differential equation for each spinor component.

We solve (\ref{etapluseqn}) in the $u$-variable defined in
(\ref{udef}).  One may use the flow equations (\ref{gravkilling}),
(\ref{chikilling2}) to show that $ M A' + M' = - \frac{1}{2}
A'^2$. One then finds
\begin{eqnarray}
\eta_+''(u) + \frac{1}{1-u} \left( \frac{5u}{2} - 1 \right) \eta_+'(u)
+ \frac{p^2 L^2}{4} \frac{1}{u(1-u)} \eta_+(u)  - && \\ 
\frac{1}{u^2 (1-u)} \left(1 - \frac{9}{8} \frac{1}{1-u} +
\frac{9}{16} \frac{(1-2u)^2}{1-u} \right) \eta_+(u) &=& 0 \,. \nonumber
\end{eqnarray}
This equation has the solution
\begin{eqnarray}
\eta_+ = u^{1/4} (1-u)^{9/4}\ 
F \left( \frac{5}{2} + \frac{1}{2} \sqrt{9 + p^2 L^2},  \frac{5}{2} -
\frac{1}{2} \sqrt{9 + p^2 L^2} ; 3 ;u \right) \eta_+^{(0)}(p)\,,
\end{eqnarray}
where $\eta_+^{(0)}(p)$ is an $r$-independent spinor.  One may read
off the spectrum: poles occur when ${5} - \sqrt{9
+ p^2 L^2} = -2n$, $n$ integer, \ie when $p^2 L^2 = 4 (n+1) (n+4)$.
This is indeed the same spectrum as the $\sigma$ scalar, as calculated
in \cite{DF} and reviewed in section 6, except for the absence of the
massless pole..

The $\eta_-$ projection is then determined by (\ref{solvetaminus}), or
alternately by deriving an equation analogous to (\ref{etapluseqn})
for $\eta_-$, by reversing the roles of $\eta_+$ and $\eta_-$
throughout.  We find the solution
\begin{eqnarray}
\label{etaminussol}
\eta_- = u^{-1/4} (1-u)^{7/4}\ F \left( \frac{3}{2} + \frac{1}{2} \sqrt{9
+ p^2 L^2},  \frac{3}{2} - \frac{1}{2} \sqrt{9 + p^2 L^2} ; 2 ;u \right) 
\eta_-^{(0)}(p) \,,
\end{eqnarray}
where only one of $\eta_\pm^{(0)}(p)$ can be specified independently.
The spectrum here is identical except the massless pole is
present in this chirality: $p^2 L^2 = 4 n(n+3)$.

The leading behavior on the boundary ($u \rightarrow 1$) is
\begin{eqnarray}
\eta_+ \sim (1-u)^{1/4} \,, \quad \quad \eta_- \sim (1-u)^{3/4} \,. 
\end{eqnarray}
Thus $\eta_+$ dominates on the boundary.  It displays the correct
scaling for a field dual to the $\Delta = 7/2$ spinor operator in the
anomaly multiplet ${\cal A}$, consistent with its limiting mass on the
boundary, $-3M(r) \rightarrow - 3/2L$.

\subsection{Coupled gravitino system}

We now discuss a solution to the coupled system of the gravitino
$\psi_\mu$ and the spin-1/2 field $\xi$, equations (\ref{psieqn}) and
(\ref{xieqn}).  Components of the $\gamma$-trace of $\psi_\mu$ mix
explicitly with $\xi$.  This is analogous to the mixing of the
graviton trace and the active scalar modulus, and that of the
longitudinal graviphoton with the active scalar phase.

The first issue is gauge fixing. Since the chiral projectors appear in
equations (\ref{psieqn}) and (\ref{xieqn}), it is useful to separate
the supersymmetry variations into the separate chiralities:   
\begin{eqnarray}
\delta \psi_{i-} &=& D_i \epsilon_- + \fracs{i}{2} A' \gamma_i \epsilon_+ \,,
\quad \quad 
\delta \psi_{i+} = D_i \epsilon_+ + \fracs{i}{2} A' \gamma_i \epsilon_- \,, \\
\delta \psi_{5-} &=& \left( D_5 + \fracs{i}{2} A' \gamma_r \right) \epsilon_-
\,, \quad \quad 
\delta \psi_{5+} = \left( D_5 + \fracs{i}{2} A' \gamma_r \right) \epsilon_+
\,, \\
\delta \xi_- &=& i \sqrt{2} \, m' \epsilon_- \,, \quad \quad 
\delta \xi_+ = 0 \,.
\end{eqnarray}
We observe that the gauge choice $\xi=0$, the analogue of the AFT
gauge of section~\ref{AnomalySec}, is not possible here since the
projection $\xi_+$ is gauge invariant. Nor can we decouple the
gravitino from (\ref{xieqn}), since both $\delta \gamma^i \psi_{i-}$
and $\delta \psi_{5-}$ depend only on $\epsilon_-$. Instead, it proves
useful to use the gauge freedom to eliminate the 4D trace
of the gravitino,
\begin{eqnarray}
\label{gauge}
\gamma^i \psi_i = 0 \,.
\end{eqnarray}
which is akin to the dynamical scalar gauge $h=0$ in the graviton
   sector. Any residual gauge transformation in this gauge must
   satisfy
\begin{eqnarray}
\delta (\gamma^i \psi_i) = \left( \Dslash_4 + 4 \fracs{i}{2} A' \right)
\epsilon = 0 \, ,
\end{eqnarray}
which is equivalent to
\begin{eqnarray}
\dslash \epsilon_- = 0 \,, \quad \quad \dslash \epsilon_+ = - 4 i A'
\epsilon_- \,.
\end{eqnarray}
Thus there are no residual gauge transformations with
   arbitrary dependence on $x^i$.

Implementing the gauge condition (\ref{gauge}), we proceed as in the
analysis of $\eta$ to decompose (\ref{xieqn}) into chiralities
\begin{eqnarray}
\label{oproj1}
i \dslash \xi_-  + 2A' \xi_+  + i \gamma^r \partial_r \xi_+ - M
\xi_+ &=& 0 \,, \\
\label{oproj2}
i \dslash \xi_+ - 2A' \xi_-
+ i \gamma^r \partial_r \xi_- - M \xi_- &=& 
- \sqrt{2} i m' \psi_{5-} \,,
\end{eqnarray}
and solve for $\xi_-$ using equation (\ref{oproj1}) to obtain
\begin{eqnarray}
\label{solvomegaminus}
\xi_- = \frac{1}{\pslash} \left( \partial_r + 2 A' - M \right) \xi_+ \,.
\end{eqnarray} 
Substituting into equation (\ref{oproj2}), we find
\begin{eqnarray}
\label{omegapluseqn}
\left(- e^{-2A} p^2 - \partial_r^2 - 5 A' \partial_r - 6 A'^2 - 2 A''
+ M^2 + M A' + M' \right) \xi_+ = - \sqrt{2} i m' \!\! \pslash \psi_{5-}  \,.
\end{eqnarray}
Thus we can solve for $\xi_+$ (and implicitly $\xi_-$)
if we can eliminate $\psi_{5-}$ from this expression.
The gravitino equation (\ref{psieqn}) allows us to do this.  In
performing the analysis, we must take care not to impose the gauge
condition (\ref{gauge}) until after all covariant derivatives have
acted, so as not to throw out nonzero terms coming from connections.
We then find
\begin{eqnarray}
\label{processedgrav}
\sqrt{2} {m'} P_+ \gamma^\mu \xi &=& i A' \delta^\mu_j ( \gamma^j \psi_5 +
\gamma_r \psi^j ) -i
\gamma^\mu \partial^j \psi_j  + i \gamma^\mu \Dslash_5
\gamma^r \psi_5 \\ &&  - i D^\mu \gamma^r \psi_5 + \fracs32  A' \gamma^\mu
\gamma^r \psi_5 + i \gamma^\mu D_5 \psi_5 + i \Dslash_5 \psi^\mu - 
\fracs32 A' \psi^\mu \,. \nonumber
\end{eqnarray}
Setting $\mu=5$, we find after a number of cancellations,
\begin{eqnarray}
- i \gamma^r \partial^j \psi_j = \sqrt{2} m' P_+ \gamma^r \xi \,,
\end{eqnarray}
leading to the chiral equations
\begin{eqnarray}
\label{partialeqns}
\partial^j \psi_{j-} = 0 \,, \quad \quad \partial^j \psi_{j+} = \sqrt{2} i m'
\xi_+ \,.
\end{eqnarray}
Furthermore, computing the $\gamma$-trace of the $\mu=i$ component 
of (\ref{processedgrav}) leads to:
\begin{eqnarray}
- 2 i \partial^j \psi_j + 3 i \Dslash_4 \gamma^r \psi_5 + 6 A'
\gamma^r \psi_5 = 4 \sqrt{2} m' \xi_- \,.
\end{eqnarray}
Taking the $P_+$ projection and using equations (\ref{partialeqns}),
we find
\begin{eqnarray}
\label{solvepsiminus}
\dslash \psi_{5-} = \frac{2 \sqrt{2}}{3} m' \xi_+ \,. 
\end{eqnarray}
The $P_-$ projection determines $\psi_{5+}$ entirely in terms of
$\xi$, as well.  

Using the relation (\ref{solvepsiminus}), we may now eliminate
$\psi_{5-}$ from (\ref{omegapluseqn}) and reach our goal, an uncoupled
ordinary differential equation for $\xi_+$.  We find
\begin{eqnarray}
\label{newomegapluseqn}
\left(- e^{-2A} p^2 - \partial_r^2 - 5 A' \partial_r - 6 A'^2 - 2 A''
+ M^2 + M A' + M'+  \frac{4}{3}  m'^2 \right) \xi_+ = 0   \,.
\end{eqnarray}
In the $u$-coordinate, this becomes
\begin{eqnarray}
\xi_+''(u) + \frac{1}{1-u} \left( \frac{5u}{2} - 1 \right) \xi_+'(u)
+ \frac{p^2 L^2}{4} \frac{1}{u(1-u)} \xi_+(u)  - && \\ 
\frac{1}{u^2 (1-u)} \left(2 - \frac{13}{8} \frac{1}{1-u} +
\frac{1}{16} \frac{(1-2u)^2}{1-u} \right) \xi_+(u) &=& 0 \,, \nonumber
\end{eqnarray}
which has the solution
\begin{eqnarray}
\label{omegaplussoln}
\xi_+(u) = u^{1/4} (1-u)^{5/4}\ F \left( \frac{3}{2} + \frac{1}{2}
\sqrt{1 + p^2 L^2}, \frac{3}{2} - \frac{1}{2} \sqrt{1 + p^2 L^2}; 3; u \right)
\xi_+^{(0)}(p)\,.
\end{eqnarray}
The spectrum here is $p^2 L^2 = 4 (n+1)(n+2)$, which agrees with other 
components of the anomaly multiplet.

We may compute $\xi_-$ from equation (\ref{solvomegaminus}).  We
obtain 
\begin{eqnarray}
\xi_-(u) = u^{-1/4} (1-u)^{7/4} \left[ 12 \ F \left( \frac{3}{2}
  + \frac{1}{2} \sqrt{1 + p^2 L^2}, \frac{3}{2} - \frac{1}{2} \sqrt{1
  + p^2 L^2}; 3; u \right) + \right. \\  
\left. u(8 - p^2 L^2)\
F \left( \frac{5}{2}
  + \frac{1}{2} \sqrt{1 + p^2 L^2}, \frac{5}{2} - \frac{1}{2} \sqrt{1
  + p^2 L^2}; 4; u \right) \right] \xi_-^{(0)}(p)  \nonumber \,.
\end{eqnarray}

The fields $\psi_{5}$ and $\partial^j \psi_j$ are then obtained
algebraically from the $\xi$ solutions.  Combined with the gauge choice
(\ref{gauge}), the only components of $\psi_\mu$ not determined 
by $\xi$ are the transverse, $\gamma$-traceless modes that we
analyzed in section \ref{TTSec}.

The behavior of the solutions near the boundary is
\begin{eqnarray}
\label{gravscal}
\xi_- \sim (1-u)^{3/4} \,, \quad \quad \xi_+ \sim (1-u)^{5/4} \, \log(1-u)
\,. 
\end{eqnarray}
In this case the negative-chirality component dominates.  It has the
correct behavior for a field dual to a $\Delta = 5/2$ operator, as
implied by its limiting mass $M(r) \rightarrow 1/2L$.

\subsection{Bianchi Identity}

As in the graviton and gauge field systems, the presence of a local
symmetry implies a Bianchi-like identity.  The supersymmetry variation
of the fermionic action must vanish:
\begin{eqnarray}
\delta_\epsilon S = \int d^5x \left( 
\delta \bar\psi_\mu \frac{\partial {\cal L}}{\partial \bar\psi_\mu} +
\delta \bar\xi \frac{\partial {\cal L}}{\partial \bar\xi} \right) =0 \,.
\end{eqnarray}
However, $\frac{\partial {\cal L}}{\partial \bar\psi_\mu} =
\frac{\partial {\cal L}}{\partial \bar\xi} = 0$ are the equations
of motion; a certain linear combination of these equations is thus
trivial owing to the gauge symmetry of the system.

Integrating by parts, we find that
\begin{eqnarray}
\delta_\epsilon S = - \int d^5x \bar\epsilon \left( (D_\mu + \fracs{i}{2}
A' \gamma_\mu )  \frac{\partial {\cal L}}{\partial \bar\psi_\mu} 
+ \sqrt{2} i m' P_+ \frac{\partial {\cal L}}{\partial \bar\xi} \right) \,.
\end{eqnarray}

Let us act on the gravitino equation (\ref{psieqn}) with the operator
$D_\mu + \fracs{i}{2} A' \gamma_\mu$, without specializing to any
particular gauge.  Note that one must be careful acting on the RHS of
(\ref{psieqn}) with $D_\mu$.  There are two terms that combine to form
a projector, $\frac{m'}{\sqrt{2}} \gamma^\mu \xi$ and $i
\frac{m'}{\sqrt{2}} \gamma^r \gamma^\mu \xi$.  The first term comes
from the Lorentz scalar $A_{abcd}$, but the second comes from $P_{5 \:
abcd} \gamma^5$.  Before acting on the second term with the covariant
derivative, we must restore it to covariant form
\begin{eqnarray}
i \frac{m'}{\sqrt{2}} \gamma^r \gamma^\mu \xi
\rightarrow i \frac{ ( \partial_\nu m) }{\sqrt{2}} \gamma^\nu \gamma^\mu \xi
\,.
\end{eqnarray}
The covariant derivative acting on $\partial_i m$ creates additional,
nonvanishing terms proportional to $m'$.

One then obtains
\begin{eqnarray}
(D_\mu + \fracs{i}{2}
A' \gamma_\mu )  \frac{\partial {\cal L}}{\partial \bar\psi_\mu} =
 - i \sqrt{2} m' P_+ \left[
i \gamma^\mu D_\mu \xi + \left( \frac{3}{2} A' + \frac{m''}{m'} \right) \xi  +
\frac{m'}{\sqrt{2}} (1+ i \gamma^r)  \gamma^i \psi_i \right] \,.
\end{eqnarray}
Using equation (\ref{Mdef}), we see this is precisely $-\sqrt{2} i m'
P_+ \frac{\partial {\cal L}}{\partial \bar\xi}$.  Thus the Bianchi identity is
satisfied; we may consider one chirality of the $\xi$ equation
as a consequence of the gravitino equation of motion.

\section{Coulomb branch fluctuations}
\label{CoulombSec}

We turn now to an examination of the $n=2$ Coulomb branch flow, as
reviewed in section \ref{ReviewCoulombSec}, an example of the class of
flow backgrounds that modify the field theory vacuum.  As discussed in
section \ref{AnomalySec}, the coupling of the graviton trace and
active scalar is universal for all RG flow backgrounds.  Here we will
examine the coupled gravitino/spin-1/2 sectors, as well as gauge
fields corresponding to unbroken $R$-symmetries.  We will find that the
fermions also have the same equations of motion as in the GPPZ case,
which is presumably a consequence of ${\cal N}=1$ SUSY.  The
graviphoton, however, behaves differently: it remains massless, as the
dual $R$-current is unbroken.  Instead, its kinetic terms are modified
by the active scalar.  We will demonstrate through field redefinition
that the modified kinetic terms are exactly equivalent to the mass
that arose in the GPPZ case.

The differences in the graviphoton sector can be traced to the fact
that the active scalar $\varphi$ is real and sits in an ${\cal N}=2$
vector multiplet.  Unlike the operator flow case, there is no phase to
play the role of a Goldstone boson --- indeed, there is no other
scalar at all in the multiplet --- and consequently the graviphoton is
left unhiggsed.  Though the $R$-current is preserved, conformal symmetry
is (spontaneously) broken; it is known \cite{SW} that in such a case
the field theory anomaly multiplet is a linear multiplet rather than a
chiral multiplet.  Hence in this case the bulk (active) vector
multiplet is dual to the operators of a linear multiplet.

The 2-point function for the TT graviton (identical to the dilaton in
this case) was first considered in \cite{FGPW2}, where it was found to
have a continuous spectrum with a mass gap $m^2_{gap} = \ell^2/L^4$.
The inert scalar $\sigma$ \cite{DF}, as well as the $h/\varphit$
system \cite{DF,AFT} which was finally solved in section
\ref{DynamicalSec}, share the continuum and gap.  All the fluctuations
we will solve for in this section have the same features.

\subsection{Truncation to {\cal N}=4 supergravity}

The equations of motion can be obtained as usual from the parent
${\cal N}=8$ theory --- an easier process than for the GPPZ flow,
since the active scalar sits in the simpler $SL(6,R)/SO(6)$
submanifold of the $E_{6(6)}/USp(8)$ scalar coset.  However, because
the large unbroken supersymmetry fully determines the equations of the
fields we are interested in, it is even easier to obtain them through
a truncation to the 5D ${\cal N}=4$ gauged supergravity theory
considered by Romans \cite{Roma}.

In the ${\cal N}=4$ theory, the gravity multiplet is the sum of ${\cal
N}=2$ gravity, gravitino and vector multiplets, and contains a
graviton, two pairs of symplectic Majorana gravitini, a single real
scalar, and gauge fields for the $SU(2) \times U(1)$ $R$-symmetry, as
well as 2-forms and two pairs of spin-1/2 fields.  The preserved
$R$-symmetry of the background is $SU(2)_L \times SU(2)_R \times U(1)_R
\subset SU(4)$, and we are free to choose an ${\cal N}=4$ subalgebra of
the full ${\cal N}=8$ that has $SU(2)_R \times U(1)_R$ as its
$R$-symmetry.  The real scalar in the ${\cal N}=4$ multiplet must be an
$R$-singlet; however, in this embedding the active scalar is the unique
real scalar with these quantum numbers.  Hence the active scalar sits
in our ${\cal N}=4$ gravity multiplet, and we can use the Lagrangian
of \cite{Roma} to read off the couplings of any fields that also
fall in this multiplet. 

In this $\cN=4$ subalgebra, the $SU(2)_L$ vectors sit in vector
multiplets and the remaining gravitini in gravitino multiplets.  We
equally well could have chosen the $SU(2)_L \times U(1)_R$ gauge
fields and the other four gravitini to sit in our ${\cal N}=4$ gravity
multiplet; all eight gravitini are equivalent in this background, as
are $SU(2)_L$ and $SU(2)_R$.  Fields such as the broken vectors,
however, cannot be placed in a massless ${\cal N}=4$ gravity multiplet
and would have to be examined by truncating the ${\cal N}=8$ theory
directly.  We shall not do so here.

To avoid confusion, note that this 5D ${\cal N}=4$ SUSY does not
contain the same 16 supercharges that are preserved in the background.
The background preserves 4D ${\cal N}=4$, which is half the
supercharges of our ${\cal N}=4$ multiplet, as well as half of the
remaining 16 generators of ${\cal N}=8$.  The 5D ${\cal N}=4$
multiplet is just a useful tool for deriving the equations of motion.
  
We have the following identifications between Romans' and our notation:
\begin{eqnarray}
\phi \longleftrightarrow \varphi \,, \qquad 
\xi \longleftrightarrow v^{1/3} \,, \qquad
g_1 \longleftrightarrow g \,, \qquad
g_2 \longleftrightarrow \sqrt{2} g \,,  \\
(\Gamma_{45})_{ab} \longleftrightarrow \delta_{ab} \,, \qquad
T_{ab} \longleftrightarrow - \frac{g}{6} W \delta_{ab} \,, 
\quad 
A_{ab} \longleftrightarrow \frac{g}{2\sqrt{2}} {\partial W \over 
\partial \varphi} \delta_{ab} 
\,.\nonumber
\end{eqnarray}

\subsection{Fermion sector}

We may now derive the equations of motion for the eight gravitini and
the eight spin-1/2 fields $\chi$ they couple to.  We combine the
fields into four sets of complex spinors as before (\ref{dirac}), and
find equations for each system that are identical to those of the GPPZ
flow:
\begin{eqnarray}
\label{psieqn2}
i \gamma^{\mu \nu \rho} D_\nu \psi_\rho &=& \frac{g}{2} W \gamma^{\mu 
\nu}
\psi_\nu + \frac{\varphi'}{\sqrt{2}} \left( 1 + i \gamma^r \right) 
\gamma^\mu \chi \,, \\
\label{chieqn2}
i \gamma^\mu D_\mu \chi &=&  M(r) \chi - \frac{\varphi'}{\sqrt{2}}
\gamma^\mu \left( 1 - i \gamma^r \right) \psi_\mu \,,
\end{eqnarray}
where again
\begin{eqnarray}
\label{Mdef2}
M(r) \equiv - \frac{3}{2} A' - \frac{\varphi''}{\varphi'} \,.
\end{eqnarray}
This universality is presumably a consequence of the ${\cal N}=1$
supersymmetry preserved in the background, despite the fact that the
5D multiplets are different in the two cases.  We conjecture that this
is generally true; it would be interesting to check whether
backgrounds with multiple scalars add new features beyond the obvious
generalization.

We have already developed the techniques to solve (\ref{psieqn2}), 
(\ref{chieqn2}) 
in section \ref{FermionSec}.  We find the uncoupled equation
\begin{eqnarray}
\label{newchipluseqn}
\left(- e^{-2A} p^2 - \partial_r^2 - 5 A' \partial_r - 6 A'^2 - 2 A''
+ M^2 + M A' + M'+  \frac{4}{3} \varphi'^2 \right) \chi_+ = 0   \,,
\end{eqnarray}
which has the solution
\begin{eqnarray}
\chi_+ = v^{a-1/6} (1-v)^{7/4} F(1+a, 2+a; 2+2a ;v) \, \chi^{(0)}_+(p) \,, 
\end{eqnarray}
where $a$ was defined below (\ref{coulombh}).  One can see that this 
solution
will have the requisite continuous spectrum and mass gap.  The
corresponding solution for $\chi_-$ is
\begin{eqnarray}
\chi_- = \frac{1}{3} \, v^{a-1/6} (1-v)^{5/4} \, [
(6a(1-v) + 2(2-5v)) F(1+a,2+a;2+2a;v) + \\ 3(a+2)v(1-v) 
F(2+a,3+a;3+2a;v) ] \, \chi^{(0)}_-(p) \,.
\nonumber
\end{eqnarray}
The behavior of the solutions near the boundary is
\begin{eqnarray}
\chi_+ \sim (1-v)^{3/4} \,, \quad \quad \chi_- \sim (1-v)^{5/4} \,
\log(1-v) \,.
\end{eqnarray}
In this case $\chi_+$ dominates on the boundary, and it scales
properly for a field dual to a $\Delta = 5/2$ operator, consistent
with its mass $M(r) \rightarrow 1/2L$.

\subsection{Vector fields}

The quadratic action for the $SU(2)_L \times SU(2)_R \times U(1)_R$
gauge fields in the Coulomb background is
\begin{eqnarray}
\label{coullag}
e^{-1} {\cal L} = - \fracs14 v^{2/3} (F^I_{\mu \nu} F^{\mu \nu I}) - 
\fracs14 
v^{-4/3} f_{\mu \nu} f^{\mu \nu} \,,
\end{eqnarray}
where $v \equiv e^{\sqrt{6} \varphi}$, 
the $F^I_{\mu \nu}, I=1 \ldots 6$ and $f_{\mu \nu}$ are field
strengths for $SO(4) \cong SU(2)_L \times SU(2)_R$ and $SO(2) \cong
U(1)_R$, respectively.  We have additionally confirmed that the Lagrangian
(\ref{coullag}) can be obtained from the ${\cal N}=8$ theory directly.
We find an equation for the transverse components
$B_k$,
\begin{eqnarray}
\label{coulveceqn}
\left( e^{-2A} \square - \partial_r^2 - 2 A' \partial_r - 
b \varphi' \partial_r \right) B_k = 0 \,,
\end{eqnarray}
where $b = 4/\sqrt{6}$ for the $SO(4)$ fields and $b=-8/\sqrt{6}$ for
the $SO(2)$.  We have used the preserved gauge symmetries to choose
$B_5 = 0$, and the field equations then require $\partial^i B_i = 0$.

The equation (\ref{coulveceqn}) resembles that for the graviphoton of
the GPPZ flow, but with the term $- b \varphi' \partial_r$ substituting
for the mass $m^2 = - 2 A''$.  Let us remove this former term through
a field redefinition $B_k \equiv \exp(-b \varphi / 2) \hat{B}_k$.  We find
\begin{eqnarray}
\label{coulveceqn2}
\left( e^{-2A} \square - \partial_r^2 - 2 A' \partial_r + m^2_B 
\right)  \hat{B}_k = 0 \,,
\end{eqnarray}
where
\begin{eqnarray}
m^2_B &\equiv& \frac{n \varphi''}{2} + b A' \varphi' + \frac{b^2 
\varphi'^2}{4} \,, \\
 &=& \frac{b}{2L^2} {\partial W \over \partial\varphi}
\left({\partial^2 W \over \partial\varphi^2} 
- \frac43 W \right) + \frac{b^2}{4 
L^2} \left({\partial W \over \partial\varphi}\right)^2 \,. 
\end{eqnarray}
For a generic choice of $W$, this need not simplify further.  However,
one may verify from equation (\ref{coulW}) that in our case
\begin{eqnarray}
\label{mysteryidentity}
{\partial^2 W \over \partial\varphi^2} - \frac{4}{3} W = 
\frac{\sqrt{6}}{3}{\partial W \over \partial\varphi} \,.
\end{eqnarray}
One may then show that precisely for $b = 4/\sqrt{6}, -8/\sqrt{6}$ 
and only for 
these values,
\begin{eqnarray}
m^2_B = - 2 A'' \,.
\end{eqnarray} 
The equation for these rescaled vectors is thus identical to that for
the transverse components of the broken graviphoton in the GPPZ
flow, and the same arguments from section \ref{TTSec} imply that their
solution is determined by the TT graviton solution, and can again be
written in terms of an auxiliary massless scalar ${f}$.

\subsection{Discussion}
\label{CoulDiscSec}

It would be interesting to understand the generality of the results we
have found coupling the gravity multiplet to the multiplet of the
active scalar.  The mixing of $h^\mu_\mu$ and $\varphit$ \cite{DF}
will be the same for any action of the form (\ref{GSaction}).  It is
natural to suspect that the form of the gravitino/spin-1/2 system,
which was identical in both backgrounds we considered, is a universal
consequence of preserving at least four supercharges.

On the other hand, the graviphoton $B_\mu$ displays two distinct
behaviors in the two examples considered, being Higgsed in the GPPZ case while
remaining massless on the Coulomb branch, although the equations of
motion turn out to be ultimately equivalent.  We are thus led to
surmise that for a general supersymmetry RG background, an active
hypermultiplet corresponds to a broken $U(1)_R$, while an active
vector (or tensor) multiplet appears when $U(1)_R$ is preserved.  As
with the breaking of conformal invariance, the coupling of the bulk
fields is not sensitive to whether the breaking is spontaneous or
explicit.

Additional support for this hypothesis comes from another
supersymmetric one-scalar flow that has appeared in the literature,
that of $\sigma$ alone \cite{GPPZ}. As we have seen this field is in a
hyper, but unlike the $m$ case, the supersymmetric flow leads to a vev
background, which preserves an $SU(3) \subset SU(4)$ but breaks
$U(1)_R$.  The fact that $m$ and $\sigma$ lead to different kinds of
SUSY backgrounds, the former an operator deformation and the latter a
shifted vacuum, is virtually invisible in the action and equations of
motion, which treats the fields symmetrically save for factors of 3;
see section \ref{InvariantSec}.  The two flows share broken $U(1)_R$,
and this breaking is accomplished in identical fashion in the two
cases through the coupling to the active hypermultiplet.  Note that it
has been suggested the $\sigma$-only flow is unphysical \cite{Gubs},
but this presumably does not affect the interplay between the
multiplet structure and the symmetries.

One is then led to suspect that the graviphoton always develops a mass
$m^2 = - 2A''$ when broken, while in the unbroken case it is instead
the kinetic terms which are modified.  Evidence in favor of such a
conjecture is that for general ${\cal N}=2$ SUGRA coupled to matter,
scalars from hypermultiplets cannot modify the vector kinetic terms,
but scalars from vector and tensor multiplets can \cite{CDA}.  We
showed that supersymmetry requires a graviphoton with canonical
kinetic terms to have $m^2 = -2A''$ in section~\ref{TTSec}, so this
obtains for all hypermultiplets.

Whether modified kinetic terms are generally equivalent to the mass term,
as they were in our case, is an important open question.
Interestingly, identities similar to (\ref{mysteryidentity}), which
was necessary to derive the equivalence, also hold for all the other
Coulomb branch superpotentials from \cite{FGPW2}.  The factor of
$\sqrt{6}/4$ is replaced by a different coefficient in each case.
However, the GPPZ flow does not possess such an identity.  It is
possible that relations like (\ref{mysteryidentity}) are somehow
characteristic of Coulomb branch flows only.  It is difficult to see
how to generalize (\ref{mysteryidentity}) to the case of multiple
scalars, however, because the $\sigma$-model indices do not seem to match
up. 

Our results for the anomaly multiplet in the case of preserved
$R$-symmetry imply that the boundary values of a 5D vector multiplet (as
well as presumably a tensor multiplet) can couple to a 4D linear
multiplet ${\cal L}_\alpha$, satisfying ${\cal D}_{\dot \alpha}{\cal
L}_{\alpha} = 0$ and ${\cal D}^{\alpha}{\cal L}_{\alpha} = \bar{\cal
D}^{\dot\alpha}\bar{\cal L}_{\dot\alpha}$, in the same way that the
gravity and hyper multiplets couple to current and chiral multiplets,
respectively.  This is not inconsistent with the expectation that a 5D
vector can couple to a 4D vector multiplet ${\cal V}$, since linear
and vector multiplets contain the same degrees of freedom: both have a
real scalar and a fermion, while the antisymmetric tensor in the
linear multiplet can be identified with the vector field possessing
the usual gauge invariance.  For a given ${\cal V}$, one may always
obtain a linear superfield,
\begin{eqnarray}
{\cal L}_{\alpha} = \bar{\cal D}^{2} {\cal D}_{\alpha} {\cal V} \,,
\end{eqnarray}
a construction familiar from the definition of the linear field strength
tensor ${\cal W}_\alpha$ in terms of a vector superfield ${\cal V}$.

\section*{Acknowledgements}

We would like to acknowledge illuminating conversations with Gianguido
Dall'Agata, Johanna Erdmenger, Stefano Kovacs, Manuel
P\'erez-Victoria, Giancarlo Rossi, Yassen Stanev, and Nick Warner.
K.P.\ would like to thank the Theoretical Physics Department of the
University of Valencia for hospitality, and D.Z.F.\ would like to
thank the Aspen Center for Physics for the same reason.  The research
of M.B. and D.Z.F. was partially supported by the
I.N.F.N.-M.I.T. ``Bruno Rossi'' exchange.\ 
The research of M.B.\ was partially supported by the EEC contract 
HPRN-CT-2000-00122 and by the INTAS project 991590.\
The research of
O.D.\ was supported by the U.S.\ Department of Energy under contract
\#DE-FC02-94ER40818 and the National Science Foundation under
PHY94-07194 and PHY99-07949.  The research of D.Z.F.\ was supported in
part by the NSF under grant number PHY-97-22072.  The research of
K.P.\ was supported in part by funds provided by the DOE under grant
DE-FG03-84ER-40168.

 \end{document}